\documentclass[graybox,vecphys]{svmult}

\usepackage{type1cm}        
\usepackage{makeidx}         
\usepackage{graphicx}        
\usepackage{multicol}        
\usepackage[bottom]{footmisc}

\usepackage{newtxtext}       %
\usepackage{newtxmath}       

\usepackage[numbers, sort&compress,comma]{natbib}
\usepackage{graphicx}
\usepackage{epsfig}
\usepackage{amsmath}
\usepackage{mathtools}
\usepackage{subfigure}
\usepackage{bm}
\usepackage{mathrsfs}
\usepackage[normalem]{ulem}

\DeclareMathOperator\erf{erf}

\def\schr{Schr\"{o}dinger equation }
\def\vr{{\bf r}}
\def\vx{{\bf x}}
\def\vR{{\bf R}}

\def\vk{{\bf k}}
\def\rr{\rho(\vr)}
\def\rrp{\rho(\vr^{ \prime})}	
\def\rzr{\rho_0(\vr)}
\def\rtr{\rho^t_0(\vr)}
\def\T{\sum\limits_{i=1}^{N}\left(-\frac{1}{2}\nabla_i^2\right)}
\def\Vext{\sum\limits_{i=1}^{N}{v_{ext}(\vr_i)}}
\def\Vee{\frac{1}{2}\sum\limits_{\substack{i,j=1 \\ i\neq j}}^{N}\frac{1}{r_{ij}}}

\def\rr{\rho(\vr)}
\def\rrp{\rho(\vr^{ \prime})}	
\def\rzr{\rho_0(\vr)}
\def\rtr{\rho^t_0(\vr)}

\def\rp{{\vr^{\prime}}}

\def \tcr{\textcolor{black}}

\def\etal{{\it et al. }}
\def\cn{C$_3$N$_4$ }

\makeindex             

\begin{document}
	
	\title*{Semiconductor Physics: A Density Functional Journey}
	\author{Sujoy Datta \thanks{Email: sujoydatta13@gmail.com} , Debnarayan Jana \thanks{Email: djphy@caluniv.ac.in}}
	\institute{Department of Physics, University of Calcutta, 92 A P C Road, Kolkata 700009, India.}
	%
	%
	\maketitle

	\abstract{The journey of theoretical study on semiconductors is reviewed in a non-conventional way. We have started with the basic introduction of Hartree-Fock method and introduce the fundamentals of Density Functional Theory (DFT). From the oldest Local Density Approximations (LDA) to the most recent developments of semi-local corrections [Generalised Gradient Approximation (GGA), Meta-GGAs], hybrid functionals and orbital dependent methodologies are discussed in detail. 
	To showcase the performance of DFT, results obtained via different approximations are compared. We indicate the success of semi-local approximations in structural properties prediction.  We also show how less computationally costly but withstand architecture of some semi-local DFT methods can solve the long riddle of bandgap underestimation. In semiconductor physics, the importance of not only the band structure prediction, but also, the proper calculation of Fermi energy, and, exact finding of band alignment is argued. 
	The comparison of Fermi energy dependent properties can channelize the theoretical studies on modern age environment-friendly researches on semiconductors, like artificial photocatalysis, energy efficient opto-electronic devices, etc. This prescription on proper choice of DFT method is potentially competent to complement the experimental findings as well as can open up a pathway of advanced semiconducting materials discoveries.}

\section{Introduction}
The journey of the many body physics in solids has never been a smooth one. The electrons to be considered are huge in number and their mutual interactions make the situation more challenging. It is computationally impossible to solve the coupled second order differential equation exactly for any solid. The Hartree-Fock (HF) antisymmetric wavefunction, which can be exactly found for any non-interacting system or for any system under averaged interaction, can not address the correlation of electrons in a real system. The Hohenberg-Kohn (HK) theorems have provided a breathing space for theorist and the era of Density Functional Theory (DFT) has begun.

The HK theorem has provided an one-to-one correspondence between the ground state charge density and the external potential. Using this, the coupled Schr\"{o}dinger equation for $N$ electrons transforms to a single particle decoupled Kohn-Sham (KS) equation. The idea of Kohn-Sham method is to find a single particle effective potential $v_{eff}(\vr)$ such that the exact ground state density $\rzr$ of the interacting system equals to the ground state density corresponding $v_{eff}(\vr)$. So, the interacting $N$ particle system is approximately transferred to a $N$ electron system $S$ under influence of averaged effective potential (\emph{S system will carry this definition throughout}). Hence, the coupled equation of $N$ interacting electron system now transforms to a single particle equation. 

Though this has been a game changer in material theory, however, different types of problem has appeared due to the approximations involved in forming $v_{eff}$.
Energy bandgap, which is the reason behind the vibrant properties of semiconductors and insulators is underestimated in most of the KS scheme based approximations due to inappropriate addressing of \tcr{exchange and correlation (xc) between electrons}. 

On he other hand, experimental investigations on semiconductors are prolific. It has started long ago, and even today, lots of effort is spent on producing more and more promising materials. From the introduction of DFT method, the theorists also try to work in harmony, but, in early days, the bandgap underestimation of primitive KS schemes made their research tremendously challenging. \tcr{An alternative method has been proposed} through the introduction of scissor operators to rigid shift the bands to match the experimental bandgaps. But this constant shift has no physical base and for those materials where no experimental bandgap data is available, this method is not useful. Furthermore, the goal of the theoretical development is not just a data matching but finding out the physical reason behind the experimental result as well as to suggest new materials with fundamental excellence and for that the solution should come from physical ground.

Perdew \etal and Sham-Schluter have provided the understanding \tcr{of the bandgap problem} in terms of discontinuity of functional derivative of xc energy with respect to (w.r.t.) electron density, namely, the derivative discontinuity \citep{perdew1982, sham1983}. Harbola and Sahni have dug into the basic quantum mechanical theory and find out the physical interpretation of this derivative discontinuity \citep{harbola1989,sahni1990}. \tcr{The next challenge has been the search for a proper approximation, which can overcome the hurdle.}


A plethora of exchange-correlation functionals and/or DFT based methods have been proposed and in most of the cases, increasing accuracy demands increasing computational cost. These are arranged in a systematic way known as Jacob's ladder \citep{perdew2001} with following rungs (i) Local Density Approximation (LDA), (ii) Generalised Gradient Approximation (GGA), (iii) Meta-GGA (MGGA), (iv) Hybrid Functional, (v) Random Phase Approximation (RPA) and beyond. The first rung is local approximation, whereas, second and third rungs are semi-local ones. \tcr{The higher rungs involve non-local terms.}

In the set of semi-local functionals, some are semi-empirical, i.e., one or more parameter of those are found by experimental data fitting, and some are non-empirical, i.e, found by fully theoretical development. Though the structural properties of materials are almost accurately predicted by some advanced non-empirical semi-local approximations, the bandgap predicted by those are highly underestimated. 
\tcr{It has been observed} that for finite systems, the accurate treatment of xc potential, particularly in the asymptotic regions, leads to eigenvalue differences between the highest occupied (HO) and lowest unoccupied (LU) orbital energies close to true excitation energies \citep{savin1998}. Most of the semi-empirical corrections intended to produce exact band gap are inspired by this observation.

Among the semi-local functionals,  vanLeeuwen-Baerends (vLB) \citep{vLB}, Becke-Johnson (BJ) \citep{becke2006} and Tran-Blaha modified BJ (TBmBJ) \citep{tran2009} type functionals are potential-only correction, which means that, they are not found by functional derivative of xc energy \citep{tran2019}.

As, HF or related schemes provide exact-exchange (EXX), so, a portion of that is mixed in various way (global, local, etc.) in hybrid functional scheme. Till date, Heyd$-$Scuseria$-$Ernzerhof (HSE) screened hybrid functional method is one the most successful and widely used method for bandgap calculation \citep{HSE, HSE1}. Almost exact band structure estimation using \tcr{this method provides} the appropriate platform for their success in calculation of optical properties of semiconductors as well.



Now, defects are ubiquitous in nature. Materials properties are highly sensitive to the formation of different types of defects. Beside the band structure, the importance of the correct determination of Fermi energy ($E_F$) in calculation of defect formation energy is also evident \citep{west2012}.
Exact calculation of $E_F$ is also a vital element in prediction of materials' work-function. Proper prediction of work-function is the key of perfect estimation of photocatalytic activity of the semiconductors. Also, the proposal of different types of semiconductor heterojunctions and metal-semiconductor Schottky junctions \citep{tung2014} rely on the exact finding of band alignment. Thus, the new era of green energy materials \citep{clark2008} becomes highly dependent on the band-alignment calculation. From three and two dimensional traditional semiconductors like silicon and graphene to relatively new materials \cn, silicene, germanene, whenever the exact prediction of work-function is needed, HSE hybrid function is a natural choice for such studies and proved to match the experimental values almost exactly \citep{west2012,datta2020carbon,singhArunima2015,datta2020pccp}.

The success of different DFT methods also depends on the choice of basis set. Results differ from pseudopotential based plane-wave basis method to site-centred methods, and, sometimes the implementation of functionals is tricky. The anomaly regarding the delocalized  HO and LU orbitals and very localized xc kernel in (semi-)local functionals can be solved by imposing the correct asymptotic behaviour of xc potential locally within a solid which can be implemented  easily in site-centred basis-set methods. That is the situation for vLB functionals.  This method which has shown success in atomic systems \citep{banerjee1999} has been implemented for the bandgap calculation of solids within localized basis of atomic sphere approximation (ASA) based linearised muffin tin orbital (LMTO-ASA) package \citep{singh2013, singh2017}.


This vLB correction to LDA has found a farm footing with the introduction of self consistent full-potential N-th order muffin tin orbital (FP-NMTO) basis sets. Using this vLB-FP-NMTO method Datta \etal have showed improvement in finding the structural properties and band gaps for orthodox group IV and group III-V materials \citep{datta2019}. This method \tcr{has further been} applied on carbon nitrides (\cn) polymorphs, and, it has helped to identify the $\gamma$-\cn as a better candidate for photocatalysis \citep{datta2020carbon}. 


We could start our discussion from the DFT theory, but there would be a gap in understanding the vastly used terms in DFT theory, like exchange-correlation, Fermi hole, self-interaction, and, so on. \tcr{Definition of those are  as much deep rooted, as the challenges of DFT approximations are.}
So, a discussion of Hartree-Fock theory is worthy to bridge the gap of understanding the difficulties of DFT approximate methods. This chapter starts with that, followed by a systematic discussion on methodological advancements in DFT till recent time. Then, we attempt to compare the performance of these DFT methods, talk about their success and failures, so that, finally this can provide a clear guideline on the proper choice of methods for investigation of \tcr{different physical} properties.

\section{The Hartree - Fock Method}\label{sec_HF}

Let us start the discussion from the time independent \schr:
\begin{equation}\label{schr}
\hat{H}\Phi=E\Phi ~; ~~ \hat{H}=\hat{T} + \hat{V}
\end{equation}
Though the kinetic energy operator $\hat{T}$ is same for all systems, however, the potential energy operator part $\hat{V}$ of the Hamiltonian $\hat{H}$ is very much system dependent. \tcr{Thus, the total energy $E$ becomes system dependent as well. The simplest two examples of electronic Hamiltonion are for Hydrogen atom and Helium atom expressed as:}
\begin{align}
	\hat{H}_H &= -\frac{\hbar^2}{2 m}\nabla^2-\frac{e^2}{4\pi\epsilon_0 }\frac{1}{r} \nonumber \\ 
	\hat{H}_{He} &= -\frac{\hbar^2}{2 m} \left(\nabla_1^2+\nabla_2^2\right)-
	\frac{e^2}{4\pi\epsilon_0 }\left(\frac{2}{r_1}+\frac{2}{r_2} - \frac{1}{r_{12}}\right) \label{eq_He}
\end{align}

Here, $m,e$ are ate electron mass and charge. \tcr{For, He, there are two electrons at $r_1$ and $r_2$ w.r.t. the centre of the ion. The Hamiltonian operator for $He$ can be decoupled (in atomic unit) as:}
\begin{equation}
\hat{T}  = - \left(\nabla_1^2+\nabla_2^2 \right) ~;~ 
\hat{V}_{ie} = \hat{v}_{ie1} +\hat{v}_{ie2} = -\left( \frac{2}{r_1}+\frac{2}{r_2} \right) ~;~
\hat{V}_{ee}= \frac{1}{r_{12}} =  \frac{1}{2|\vr_1-\vr_2|} \label{schr_He}
\end{equation}

The  ion-electron (\emph{i-e}) interaction operators ($\hat{v}_{ie1}, \hat{v}_{ie2}$) are operating on single electrons, and have the dimension of potential, whereas, electron-electron ($ee$) interaction operator $\hat{V}_{ee}$ is a two-body energy operator, \tcr{which depends on the relative distance between the electrons}. We write potential operators in small case and energy operators in capital throughout the chapter.

Due to the \tcr{$\hat{V}_{ee}$ term}, the \schr for two electrons can not be decoupled into two. \tcr{As the atomic number increases, it becomes tougher to solve the coupled equation analytically.} In addition, for molecules and condensed matter systems, there are more than one ions (atomic core) involved, so there should be ion-ion interaction as well. As the ions are heavy, they move very slowly than the electrons, so, we can begin by separating the `fast' electronic from the `slow' ionic degrees of freedom. This is the adiabatic or the  {\bf Born-Oppenheimer approximation} \citep{born1927}.

Situation would be simple if $\hat{V}_{ee}$ could be written in a summative form over these two electrons, so that, we could write an effective potential $\hat{v}_{s}$ replacing $\hat{v}_{ie}+\hat{v}_{ee}$. This is an {\bf independent particle model} followed by {\bf Slater-Hartree method} and then $\Phi (\vR) =\phi_1(\vr_1) \phi_2(\vr_2)$, where, $\vR=\{\vr_1,\vr_2\}$. Now, according to the Pauli exclusion principle, the wavefunction should be anti-symmetric, so, adding Fock's contribution the {\bf Hartree-Fock} wavefunction becomes: $ \Phi^{HF}(\vR)=
\begin{vmatrix}
\phi_1(\vr_1) & \phi_2(\vr_1) \\
\phi_1(\vr_2) & \phi_2(\vr_2) 
\end{vmatrix}$.
This is a $2$ electron Slater type wavefunction build up out of single electron orbitals for a system where each electron moves in an effective potential including average effect of $ee$ repulsion.

Going back to the Eq. [\ref{schr_He}], except $\hat{V}_{ee}$, \tcr{all other terms can be expressed as sum over the number of electrons, i.e., they are independent-electron term. Hence, the contribution of these terms in the total energy are also additive over the electron number $N$. The remaining \emph{e-e} interaction energy term can be expressed as:}
\begin{align}
&E^{HF}_{ee} = \langle \Phi^{HF}| \hat{V}_{ee} |\Phi^{HF}\rangle \nonumber \\
=   \iint \frac{d\vr_1 d\vr_2}{2|\vr_1-\vr_2|} &
\left[ \phi^\ast_1(\vr_1)\phi^\ast_2(\vr_2) \{\phi_1(\vr_1)\phi_2(\vr_2) - \phi_2(\vr_1)\phi_1(\vr_2) \} \right. \nonumber \\
&\left.- \phi^\ast_2(\vr_1)\phi^\ast_1(\vr_2)\{\phi_1(\vr_1)\phi_2(\vr_2) -\phi_2(\vr_1)\phi_1(\vr_2) \} \right] \nonumber \\
=  \iint \frac{d\vr_1 d\vr_2}{2|\vr_1-\vr_2|} &
\{|\phi_1(\vr_1)|^2 |\phi_2(\vr_2)|^2 + |\phi_2(\vr_1)|^2 |\phi_1(\vr_2)|^2\} \label{v_ee_HF} \nonumber\\ 
-  \iint \frac{ d\vr_1 d\vr_2}{2|\vr_1-\vr_2|} &
\{ \phi^\ast_1(\vr_1)\phi_1(\vr_2)\phi^\ast_2(\vr_2) \phi_2(\vr_1)
- \phi^\ast_1(\vr_2)\phi_1(\vr_1)\phi^\ast_2(\vr_1)\phi_2(\vr_2)\}   
\end{align}

See, $\phi_1$ at $\vr_1$ and $\vr_2$, i.e., $\phi_1(\vr_1)$ and $\phi_1(\vr_2)$ are same electronic state at two \tcr{positions}, so, $\phi_1(\vr_1)\phi_1(\vr_2)$ type term represents $e-e$ self-interaction, which is unphysical.
In the terms of Eq. [\ref{v_ee_HF}], self-interaction term, i.e., 
$|\phi_1(\vr_1)|^2 |\phi_1(\vr_2)|^2 + |\phi_2(\vr_1)|^2 |\phi_2(\vr_2)|^2$ is absent, so, \textbf{\emph{HF theory is a theory of exact-exchange and is self-interaction corrected (SIC)}} from the very beginning.

\subsection{Hartree - Fock Method for $N$ electron system}\label{sec_HF_N}

Let us now generalize the idea to electronic many-body systems beyond He atom. \tcr{The general time-independent \schr involving many-electron Hamiltonian is:}  
\begin{align}
\hat{H}\Psi({\bf X}) &= E\Psi({\bf X}) \text{  ; } {\bf X} = \{ \vr_1 \bm{\sigma}_1, \vr_2\bm{\sigma}_2,.., \vr_N\bm{\sigma}_N\}
    ~~ \& ~~~ \hat{H}=\hat{T}+\hat{V}_{ext}+\hat{V}_{ee} \label{Hamiltonian}\\
\text{where,}~~&\hat{T}=\T \text{  ;  } \hat{V}_{ext} = \Vext \text{  ;  } \hat{V}_{ee}=\Vee 
\end{align}

This $v_{ext}(\vr_i)$ is the external potential which includes the ion-electron ($ie$) interaction, \tcr{and, local in nature}.  \footnote[1]{An operator $\hat{A}$ is local if it follows $\hat{A}(\vr,\vr')=\hat{A}(\vr')\delta(\vr-\vr')$. Simply, it means that, the operator acting at any point $\vr$ does not depend on anything at $\vr'$. Potential part of one-body Hamiltonian is often local.}
The many-electron wave-function is a function of combination of coordinates ($\vR=\{\vr_i\} = \{\vr_1, \vr_2, ..., \vr_N\}$) and spins ({\bf $\sigma$}$=\{\sigma_i\} = \{\sigma_1, \sigma_2, ..., \sigma_N\}$) of individual electrons.
The normalized spin-orbital wavefunction of HF system $\Psi^{HF}({\bf X})$ is of single Slater determinant ($\Psi^{S}$) form:
\begin{align}\label{wf_HF}
\Psi^{HF}({\bf X}) \equiv \Psi^{S}({\bf X})= \frac{1}{\sqrt{N!}} 
\begin{vmatrix}
\psi_1({\bf x}_1) & \psi_2({\bf x}_1) & \cdots & \psi_N({\bf x}_1)\\
\psi_1({\bf x}_2) & \psi_2({\bf x}_2) & \cdots & \psi_N({\bf x}_2)\\
\vdots & \vdots & \ddots & \vdots \\
\psi_1({\bf x}_N) & \psi_2({\bf x}_N) & \cdots & \psi_N({\bf x}_N)
\end{vmatrix}  ;~ {\bf x_n}= {\vr_n \sigma_n}
\end{align}

Here, $\frac{1}{\sqrt{N!}}$ is the normalisation constant and the spin-orbitals follow orthonormality condition: $\int \psi^{\ast}_i(\vx_n) \psi_j(\vx_n) d\vx_n =\delta_{ij}$. Since spin-orbitals corresponding spin up states are automatically orthonormal to those of spin down states, so, this orthonormality condition reduces to the orthonormality of space-only orbitals ($\{\phi_i(\vr_n)\}$) having parallel spin.
Writing the spin-orbitals $\psi_i(\vx_n) = \phi_i(\vr_n)\chi_i(\sigma_n)$ and as the spin functions are orthonormal, we can build the spatial only HF wavefunction $\Phi^{HF}({\bf R})$ through replacing  $\{\psi_j({\bf x}_n)\}$ by $\{\phi_j({\bf r}_n)\}$. 


Note that, the Slater type wavefunction requires separation of variable for each spin-orbitals which is only possible if there is no two-body term in the Hamiltonian. Two options are there, either $\hat{V}_{ee}=0$, or, $\hat{V}_{ee}$ is approximated as an averaged potential. In HF scheme, we try to build a Slater type wavefunction for system having $\hat{V}_{ee} \ne 0$. Using this trial wavefunction $\Psi^{HF}({\bf X}) \equiv \Psi^{S}({\bf X})$, let us express the \emph{e-e} interaction energy $E_{ee}$ \citep{hartree1928, parr1980}.

\begin{align}
E_{ee}^{HF} = \langle \Psi^{HF}| \hat{V}_{ee} |\Psi^{HF}\rangle
= \frac{1}{2} &\sum_{i,j=1}^{N}\left( J_{ij}-K_{ij} \right)  = E_H + E_x \label{E_ee_HF}\\
J_{ij} =  \iint \frac{\psi_i^{\ast}(\vx) \psi_i(\vx) \psi_j^{\ast}(\vx') \psi_j(\vx') }{|\vr-\vr'|} d\vx d\vx' &;
K_{ij} =  \iint  \frac{\psi_i^{\ast}(\vx) \psi_i(\vx') \psi_j^{\ast}(\vx') \psi_j(\vx) }{|\vr-\vr'|} d\vx d\vx' \nonumber
\end{align}

Here, $J_{ij}$ and $K_{ij}$ are  {\bf Coulomb interaction integral} and {\bf exchange energy integral} and the sums are over all $\{i,j\}$, as the self terms: $J_{ii}=K_{ii}$ cancel each other.

The  {\bf Hartree energy} $(E_H)$ and corresponding {\bf Hartree potential} $v_H(\vr)$ are expressed by summing over $J_{ij}$ through the definition of density $\rho(\vr) = \sum\limits_{i=1}^N \psi^\ast_i(\vx) \psi_i(\vx)$ as:
\begin{align}
E_H = \frac{1}{2} \sum_{i,j=1}^N J_{ij}=  \frac{1}{2} \iint  \frac{\rr \rrp}{|\vr-\vr'|} d\vr d\vr' ~;~~
v_H(\vr) = \int  \frac{\rrp}{|\vr-\vr'|} d\vr' \label{v_H} 
\end{align} 

\begin{svgraybox}
\textbf{\large B1: Electron Density and Density Matrix}

The number of electrons per unit volume at a position $\vr_1$ in a given state is defined as the electron density for that state at $\vr_1$:
\begin{equation}
\rho(\vr_1) = N \idotsint \Psi^\ast({\bf X}) \Psi({\bf X}) d\sigma_1 d\vx_2 \cdots d\vx_N ~;~ \int \rho(\vr_1) d\vr_1 = N
\end{equation}

For Slater type wavefunction $\Psi^S({\bf X})$ (as in HF scheme)
\begin{align}
\int [ \cdots] d\vx \equiv \sum_\sigma \int [ \cdots] d\vr \Rightarrow
\rr= \sum_\sigma \sum_{i}^{occ}|\psi_i(\vr_i \sigma)|^2 
\end{align}

The { \bf reduced density matrices} of first and second order are:
\begin{equation}
\begin{aligned}
&\gamma_1(\vx_1,\vx_1') = N \int  \Psi^\ast(\vx_1, \vx_2, \vx_3,\cdots, \vx_N ) \Psi(\vx_1', \vx_2, \vx_3,\cdots, \vx_N )  d\vx_2 \cdots d\vx_N \\
&\gamma_2(\vx_1\vx_2,\vx_1'\vx_2') = {N \choose 2} \int \Psi^\ast(\vx_1, \vx_2, \vx_3,\cdots, \vx_N ) \Psi(\vx_1', \vx_2', \vx_3,\cdots, \vx_N )  d\vx_3 \cdots d\vx_N 
\end{aligned}\label{gamma_1}
\end{equation}

The first-order and second-order {\bf spinless reduced density matrices} are:
\begin{equation}
\begin{aligned}
\rho_1(\vr_1,\vr_1') &= \int \gamma_1(\vr_1\sigma_1,\vr_1'\sigma_1) d\sigma_1 
\Rightarrow \rho(\vr_1) = \rho_1(\vr_1,\vr_1) \label{rho_1}\\
\rho_2(\vr_1\vr_2,\vr_1'\vr_2') &= \iint \gamma_2 (\vr_1\sigma_1 \vr_2\sigma_2, \vr_1'\sigma_1 \vr_2'\sigma_2)
d\sigma_1 d\sigma_2 
\end{aligned}
\end{equation}
\begin{equation}
\text{For Slater wavefn.}~ \rho_1(\vr_1,\vr_1') = \sum_{\sigma_1} \sum_{i}^{occ}  \psi_i^{\ast}(\vr_1\sigma_1) \psi_i(\vr_1'\sigma_1) \label{density_spin}
\end{equation}

Here, $\vr_1, \vr_1', \sigma_1$ are dummy index, so, can simply be replaced by $\vr, \vr', \sigma$. $\sum\limits_{\sigma_1}$ is for up and down spins,  $\sum\limits_{i}^{occ}$ is for all occupied orbitals and both sum yields total $N$ number of sum over $N$ spin-orbitals.
For closed shell, every orbital is doubly spin degenerate. As a result, the summation over $i$ runs from $1$ to $N/2$ occupies orbitals. Another thing to notice is that, in spinless reduced density matrices, integrations are done over all spins, $\sigma_2 \cdots \sigma_N$ in Eq. [\ref{gamma_1}] and on $\sigma_1$ in Eq. [\ref{rho_1}], so, we write in term of only space orbitals as:
\begin{equation}
\rho(\vr)= 2 \sum_{i}^{N/2}|\phi_i(\vr)|^2 =  \rho_\uparrow (\vr) + \rho_\downarrow (\vr) 
~~\&~~ \rho_1(\vr,\vr') = 2 \sum_{i}^{N/2}  \phi_i^{\ast}(\vr) \phi_i(\vr') \label{density}
\end{equation}
\end{svgraybox}

\subsection{Hartree - Fock Equation}
If we want to use the HF scheme for the interacting electronic system, then, we have to start with taking a trial wavefunction $\Psi({\bf X})$ of Slater determinant type (Eq. [\ref{wf_HF}]).
Using the ortho-normalization condition at any position $\vr$, $\int \psi^{\ast}_i(\vx) \psi_j(\vx) d\vx =\delta_{ij}$ and $\frac{\partial \langle \hat{H} \rangle}{\partial \psi^{\ast}_i(\vx)}=0$, we reach to {\bf HF equation} applying Lagrange's multiplier method:
\begin{align}\label{HFeq_s}
\left(-\frac{1}{2}\nabla_i^2\right) \psi_i(\vx) + v_{ext}(\vr)\psi_i(\vx) 
&+ \left[\sum\limits_{j=1}^{N} \int  \frac{\psi_j^{\ast}(\vx') \psi_j(\vx')} {|\vr-\rp|} d\vx'\right] \psi_i(\vx) \nonumber\\
&- \left[\sum\limits_{j=1}^{N}  
 \int  \frac{\psi_j^{\ast}(\vx') \psi_i(\vx')} {|\vr-\rp|} d\vx'\right] \psi_j(\vx) = \epsilon_i \psi_i(\vx)
\end{align}

For a closed shell system having even number of electrons the $N$ spin-orbitals can be divided into $N/2$ space-only orbitals, and for each of these, two spin (up and down) states.

\tcr{The {\bf Kinetic energy} contribution to the total energy can be calculated exactly as:}
\begin{align}
T =\langle \Psi^{HF}| \hat{T} |\Psi^{HF}\rangle 
= - \frac{1}{2} \sum_{\sigma} \sum_i\int \psi^{\ast}_i(\vr, \sigma)\nabla_i^2 \psi_i(\vr, \sigma) d\vr
\end{align}

If we take inner product of HF Eq. [\ref{HFeq_s}] with $\psi_i(\vx)$ and sum over $\{i\}$ then the {\bf HF total energy} $E[\Psi^{HF}]$ in terms of orbital energy $\epsilon_i$ is found using Eq. [\ref{E_ee_HF}] as:
\begin{align}
\langle \Psi^{HF}| \hat{T}+\hat{V}_{ext} |\Psi^{HF}\rangle &+ 2 E_{ee}^{HF} =\sum_{i}^{N} \epsilon_i \nonumber\\
\Rightarrow E[\Psi^{HF}] = \langle \Psi^{HF}| \hat{T} &+\hat{V}_{ext} |\Psi^{HF}\rangle + E_{ee}^{HF} =\sum_{i}^{N} \epsilon_i - E_{ee}^{HF}
\end{align}

From the last term of Eq. [\ref{HFeq_s}] the {\bf orbital dependent exchange operator} $\hat{v}_{x,i}(\vr)$ and {\bf orbital dependent Fermi hole} $\rho_{x,i}(\vr,\rp)$ are defined as\citep{harbola1989, sahni1990}:
\begin{align}
&\hat{v}_{x,i}(\vx) \psi_i(\vx) 
= -  \left[\sum\limits_{j=1}^{N} \int  
\frac{\psi_j^{\ast}(\vx') \psi_i(\vx') \psi_j(\vx)} {\psi_i(\vx)|\vr-\rp|} d\vx'\right] \psi_i(\vx) 
= \left[\int \frac{\rho_{x,i}(\vr,\rp)}{|\vr-\rp|} d\vr' \right] \psi_i(\vx) \label{XP} \\
&\text{Using}~~\int d\vx \equiv \sum_{\sigma} \int d\vr ~;~~ \vx = \vr \sigma, \vx' = \vr' \sigma ~~\text{we find:} \nonumber\\
& \rho_{x,i}(\vr,\rp) = - \frac{\sum\limits_\sigma \sum\limits_{j}^{occ} \psi_j^{\ast}(\vx') \psi_i(\vx') \psi_j(\vx)} {\psi_i(\vx)} 
= - \sum_\sigma \sum_{j} \frac{ \psi_i^{\ast}(\vx) \psi_j^{\ast}(\vx') \psi_i(\vx') \psi_j(\vx)} 
{ \psi_i^{\ast}(\vx) \psi_i(\vx)} \label{orb_Fermi_hole}\\
&~\text{  and,  }~ \int \rho_{x,i}(\vr,\rp) d\rp = -1
\end{align}

The double sum yields $N$ electronic spin-orbitals with sum over $\{\sigma\}$ running over up$+$down spins and sum over $\{i\}$ over all occupied orbitals corresponding each type of spin.
\tcr{Eq. [\ref{XP}] shows that the exchange potential operator $\hat{v}_{x,i}$ is not multiplicative, i.e., {\bf non-local operator}. Furthermore, due to the \emph{dynamical}, i.e., orbital dependent nature of hole, charge distribution $\rho_{x,i}$, and, the corresponding potentials are orbital dependent.} Therefore, \emph{ \textbf{HF equation and theory is orbital dependent}}, and, the calculation becomes rigorous. 


\subsection{Hartree - Fock - Slater Equation} \label{sec_HFS}
Till now, \tcr{we have not talked about any approximation within the HF theory}. 
Let us note the condition, $\int \rho_{x,i}(\vr,\rp) d\rp=-1$ is true for each electron (occupying an spin-orbital $\psi_i$) at position $\vr$. So, an weighted averaging of $\rho_{x,i}(\vr,\rp)$ over all orbitals can provide the expression of the orbital-independent, or, simple {\bf Fermi hole} $\rho_x$.
\begin{align}
\rho_x (\vr, \vr')= 
\sum_\sigma \sum_{i}^{occ} \rho_{x,i}(\vr,\rp) p_i(\vr)
=-\left[ \frac{\sum_\sigma\sum_{i,j}^{occ} \psi_i^{\ast}(\vx) \psi_j^{\ast}(\vx') \psi_i(\vx') \psi_j(\vx)} 
{\sum_\sigma\sum_{k} \psi^{\ast}_k(\vx) \psi_k(\vx)} \right]  \label{Fermi_hole}\\ 
\text{where,}~ p_i(\vr) = \frac{\psi^{\ast}_i(\vx) \psi_i(\vx)} 
{\sum\limits_\sigma\sum\limits_{k}^{occ} \psi^{\ast}_k(\vx) \psi_k(\vx)} ~\text{is the probability}~\&
~  \int \rho_{x}(\vr,\rp) d\rp = -1
\end{align}

Now, for even $N$ numbered electronic system, forming closed shells, we can define the Fermi hole using the spinless reduced density matrix using Eq. [\ref{density_spin}] and [\ref{density}] as:
\begin{align}
\rho_x (\vr, \vr') = - \frac{|\rho_1 (\vr, \vr')|^2/4}{\rho(\vr)/2} = -\frac{|\rho_1 (\vr, \vr')|^2}{2\rho(\vr)} &
~~~~~~~ \textbf{[closed shell]} \label{ex_density1}
\end{align}

If we approximate $\rho_{x,i}(\vr,\rp)$ of Eq. [\ref{XP}] by $\rho_{x}(\vr,\rp)$ then, the exchange potential operator $\hat{v}_{x,i}(\vr)$ becomes {\bf multiplicative or local operator} $v_{x}^{S}(\vr)$ and the {\bf Hartree-Fock-Slater equation} from Eq. [\ref{HFeq_s}] using Eq. [\ref{v_H}] becomes:
\begin{equation}\label{S-HFeq}
\left[ -\frac{1}{2}\nabla_i^2 + v_{ext}(\vr) + v_H(\vr) + v_{x}^{S}(\vr) \right] \psi_i(\vx) = \epsilon_i \psi_i(\vx)
\end{equation}
Where, the Slater potential $v^S_x(\vr)$ is defined in term of Fermi hole and  {\bf exchange only pair-correlation function} $g_{x}(\vr,\vr')$ as:
\begin{equation}
v_{x}^{S}=\int \frac{\rho_{x}(\vr,\rp)}{|\vr-\rp|} d\vr' = \int \frac{\rrp \{g_{x}(\vr,\rp) - 1\}}{|\vr-\rp|} d\vr'
~;~  g_{x}(\vr,\vr') = 1 + \frac{\rho_{x}(\vr,\vr')}{\rrp } \label{g_x}
\end{equation}
Using $\rho_{x}(\vr,\rp)$, we can also define the {\bf Pauli exchange energy} $E_x^{HF}$ using Eq. [\ref{E_ee_HF}]:
\begin{equation} \label{E_x_HF}
E_x^{HF} = -\frac{1}{2} \sum_{i,j=1}^N K_{ij}= \frac{1}{2} \iint  \frac{\rr \rho_{x}(\vr,\rp)}{|\vr-\vr'|}  d\vr d\vr'
\end{equation}

This is how a many electron \schr is developed in Slater-Hartree-Fock theory without any two-body term in Hamiltonian. As, electronic many body wavefunction acting under averaged effective potential can be written as Slater determinant, starting the calculation with $\Psi^S$ is justified. As for the system, the Fermi hole is developed due to the self interaction and Pauli exchange effect, the exchange-only interpretation of Fermi hole $\rho_x$ is justified as well.

An interesting system is a system of (spin polarised) single electron. For that no sum is needed in Eq. [\ref{Fermi_hole}] and $i,j,k,\sigma =1$ in the definition of Fermi hole and density. So, for this special case $\left.\rho_{x}(\vr,\rp)\right|_1 = - \left.\rho (\vr')\right|_1$ and the exchange energy $\left.E_x^{HF}\right|_1 = - \left. E_H\right|_1 \Rightarrow \left.E_{ee}^{HF}\right|_1 = 0$, which implies:

\begin{center}
	 \emph{\textbf{ Hartree-Fock theory is single electron self-interaction free.}}
\end{center}

\vskip 0.1cm
{\bf NOTE:}
If the wavefunction $\Psi({\bf X})$ is found for interacting system, i.e, as a solution of $\hat{H}$ including $\hat{V}_{ee}$, then the coulomb correlation effect also counts in.
For such interacting electronic system, {\bf pair-correlation function} $g_{xc}(\vr,\vr')$ and {\bf Fermi-Coulomb hole} $\rho_{xc}(\vr,\vr')$ includes Coulomb interaction between electrons.
\begin{align}
\rrp ~ g_{xc}(\vr,\vr') &= \rrp + \rho_{xc}(\vr,\vr') ~;~~ \int  \rho_{xc}(\vr,\vr')d\vr' = -1 \label{g_xc}\\
\therefore E_{ee}
&=  \frac{1}{2} \iint  \frac{\rr \rrp g_{xc}(\vr,\vr')}{|\vr-\vr'|} d\vr  d\vr' \nonumber\\
&= \underbrace{\frac{1}{2} \iint  \frac{\rr \rrp}{|\vr-\vr'|} d\vr  d\vr'}_{\substack{\textbf{Hartree energy ($E_H$)}}}   
+ \underbrace{\frac{1}{2} \iint  \frac{\rr \rho_{xc} (\vr,\vr')}{|\vr-\vr'|} d\vr  d\vr'}_{\substack{\textbf{Pauli-Coulomb energy ($E_{xc}$)}}} \label{E_H}
\end{align}

\section{Hohenberg-Kohn (HK) Theorem} \label{sec_HK}
\begin{itemize}
	\item {\bf First theorem:}
	{\emph {The non-degenerate ground state density $\rzr$ determines  the external potential $v_{ext}(\vr)$ to within a trivial additive constant.}}
	\item {\bf Second Theorem:}
	{\emph { The non-degenerate ground state density $\rzr$ can be determined from the ground state energy functional $E[\rho_0]$ via the variational principle by varying only the density.}}
\end{itemize}

The ground state non-degenerate density $\rzr$ exactly determines the electron number $N$, the ground state wave-function $\Psi[\rzr]$ and all other electronic properties. \tcr{Not necessarily, $\Psi$ is a single Slater type wavefunction $\Psi^{S}$ of Eq. [\ref{wf_HF}], but, can be any general $N$ electron wavefunction.} In addition according to the first HK theorem, $\rzr$ determines $v_{ext}(\vr)$ as well \citep{hohenberg1964, kohn1965}. So, what does it mean? We can solve $N$-electron \schr having $\hat{V}_{ext}(\vr)$ in Hamiltonian to find the ground state wavefunction $\Psi$ (map $A$: $\hat{V}_{ext}(\vr) \rightarrow \Psi$), and then, use $\Psi$ to find the ground state density $\rzr$ (map $B$: $\Psi \rightarrow \rzr$). This $AB$ mapping is what the first theorem provides. It also provides the definition of $v$-representability. \footnote[2]{If we solve \schr with any $v_{ext}(\vr)$ to find the antisymmetric ground state wave function $\Psi$ and using this we find $\rzr$, then $\rzr$ is \textbf{v-representable}.}

Now the first question is the inverse mapping $(AB)^{-1}$ exists? The answer is \emph{yes}, and, the existence can be proved easily \citep{parr1980, sahni2016}. Second question is how this can be found? The $A^{-1}$ can be found using the inverse of the \schr to within a constant. Another way is to use {\bf quantal Newtonian first law} using force filed. This is where the {\bf Quantal DFT (Q-DFT)} starts to diverge, and detail study can be found in \citep{sahni2016}. As, there can be infinite number of antisymmetric wavefunctions that integrate to form the ground state density $\rzr$, the inverse mapping $B^{-1}$ is not that straightforward as $A^{-1}$, and can be found using {\bf constrained search} approach \citep{levy1985}. 

Let, $\rtr$ is the trial ground state density.
According to first theorem, this $\rtr$ exactly determines the trial wave-function $\Psi^t[\rho^t_0]$ and external potential $v^t_{ext}(\vr)$.
\begin{equation}\label{eq_HK2}
\text{Now,}~~ E^t=E[\rho^t_0] = \langle \Psi^t|\hat{H}|\Psi^t\rangle \geqslant E \text{  (ground state energy)}
\end{equation}
As, no other variable is involved in the process of this equation except the density, so, variational principle can be applied using $\int \rzr d\vr =N$ as:
\begin{equation}\label{Euler}
\delta \{E[\rho_0] - \mu(\int \rzr d\vr -N) \}=0 ~~~~~~~ \Rightarrow \frac{\delta E[\rho_0]}{\delta \rzr}=\mu 
\end{equation}

This is the {\bf Euler-Lagrange equation}.
Let, $\rho^{(N)}_0(\vr)$ is the density (solution of Eq. [\ref{Euler}]) for $N$ electron system with ground state energy $E[\rho^{(N)}_0]$. If the charge is changed as $N \rightarrow N+f$, then the energy difference is:
\begin{align}
E^{(N+f)}& - E^{(N)} =
 \int \left. \frac{\delta E[\rho]}{\delta \rr} \right|_{\rho^{(N)}_0} \{\rho^{(N+f)}_0(\vr) -\rho^{(N)}_0(\vr)\}d\vr \nonumber\\
&=\mu(N) \int \{\rho^{(N+f)}_0(\vr) -\rho^{(N)}_0(\vr)\}d\vr 
= \mu(N)\{(N+f)-N\} = f \phantom{5}\mu(N) \\
&\text{For } f \to 0: \lim_{f \to 0} \frac{E^{(N+f)}-E^{(N)}}{(N+f)-N} =\frac{\partial E[N]}{\partial N} =\mu^{(N)} \label{mu_N}
\end{align}
So, $\mu^{(N)}$ is the energy to change the electron number, so, $\mu^{(N)}$ represents the {\bf chemical potential} for $N$ electron system.

The {\bf ionization energy} ($I$) for removing one electron from $N$ electron system and {\bf electron affinity} ($A$) for adding one electron in that are, respectively:
\begin{equation} \label{mu_IA}
	\mu^{(N)}_{-1} = \frac{E^{(N-1)}-E^{(N)}}{(N-1)-N} =-I^{(N)} ~~~\&~~~ \mu^{(N)}_{+1} = \frac{E^{(N+1)}-E^{(N)}}{(N+1)-N} =-A^{(N)}
\end{equation}
\tcr{Consequently,} the {\bf fundamental (Optical) band gap $E_g$ for isolated system} of $N$ electron is:
\begin{equation}\label{E_g}
E_g = I^{(N)}-A^{(N)} = \mu^{(N)}_{+1}- \mu^{(N)}_{-1}
\end{equation}

In the Hamiltonian operator, $\hat{T}$ and $\hat{V}_{ee}$ have same form for every many electronic system. The external potential energy operator $\hat{V}_{ext} = \hat{V}_{ext} [\rho]$ is system dependent, as it depends on the external potential, i.e., knowledge of nuclei or other sources have to be known.
So, we can decompose the total energy as:
\begin{align}\label{E_HK}
E[\rho_0] = \langle \Psi[\rho_0] | \hat{H} | \Psi[\rho_0]\rangle =& F^{HK} [\rho_0] + \int \rzr v_{ext} (\vr) d\vr \\
\text{where,}~~ F^{HK}[\rho_0] =& \langle \Psi[\rho_0] | (\hat{T} +\hat{V}_{ee}) | \Psi[\rho_0]\rangle \label{F_HK}
\end{align}
This $F^{HK}[\rho_0]$ is the {\bf universal functional of density}, which is same for all electronic system.
Using Eq. [\ref{E_HK}], the {\bf Euler-Lagrange Equation} [\ref{Euler}] becomes :
\begin{equation}\label{Euler_N}
 \frac{\delta F^{HK}[\rho_0]}{\delta \rzr} + v_{ext} (\vr) =\mu 
\end{equation}

\begin{figure}[]
		\sidecaption
	\framebox{\includegraphics[scale=0.22]{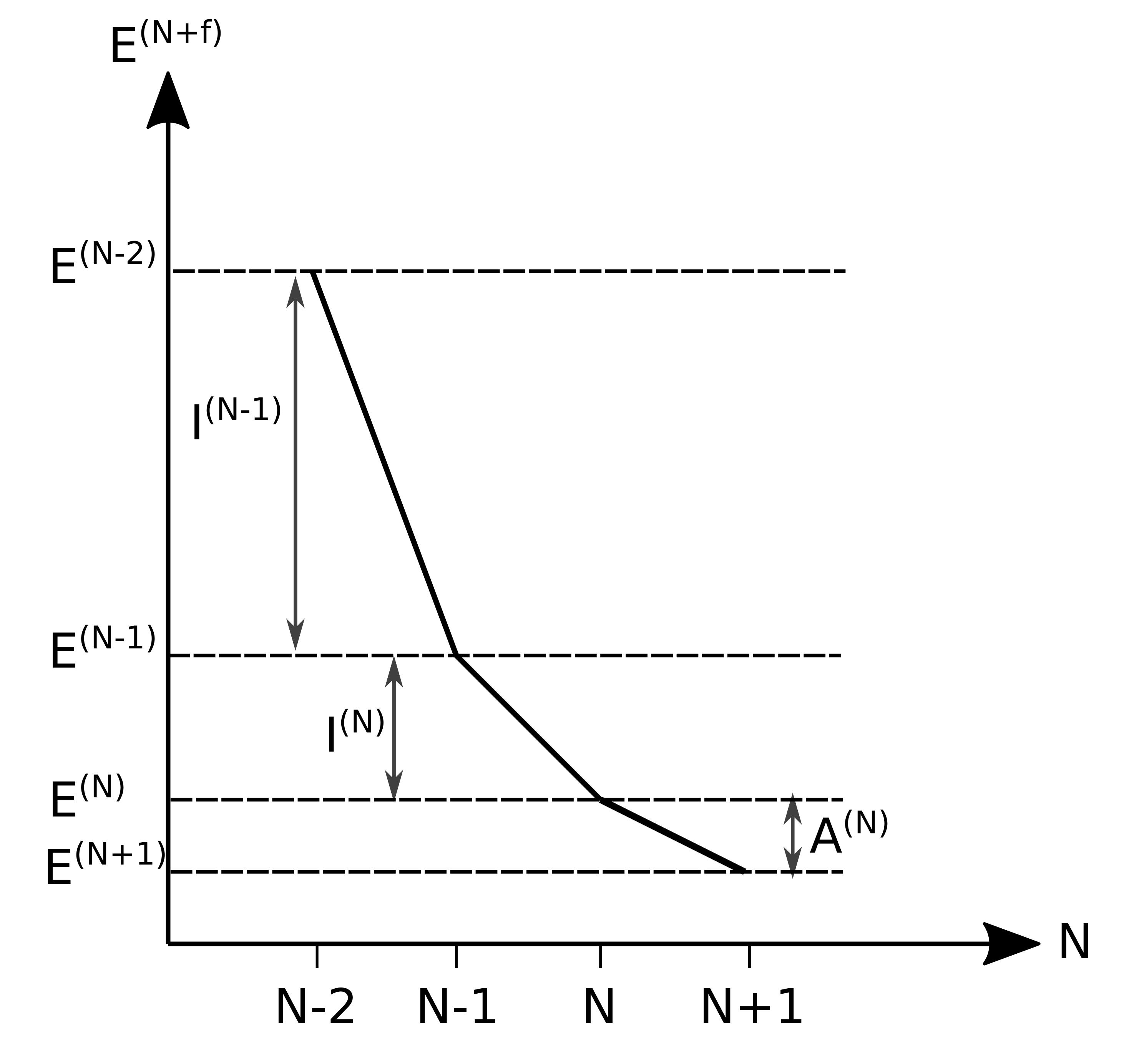}}
	\caption{Variation of energy $E^{(N+f)}$ with fractional electron number $N$ showing the piecewise linearity. $I^{(N-1)} > I^{(N)}$ is also evident.}
	\label{fig1}       
\end{figure}

\begin{svgraybox}\label{box2}
\textbf{\large B2: Fractional Electron Number and Derivative Discontinuity}

The derivation of Eq. [\ref{mu_N}] mathematically uses a condition of fractional electron number $N$. In an open system where electron exchange is possible between atoms, the time average of charge can be fractional. This has been shown by Perdew, Parr, Levy and Balduz (PPLB) \citep{perdew1982}. The extension to fractional spins and the further combination of fractional charges and spins have been done later \citep{cohen2008}. As quantum mechanical systems with fluctuating number of particles are described by ensemble mixture, using Levy's constrained search formalism \citep{levy1985} the charge density, energy and chemical potentials are found as:
\begin{equation}
\rho_0^{(N+f)} = (1-f) \rho_0^{(N)} +f\rho_0^{(N+1)} ~~\&~~~
E^{(N+f)} = (1-f) E^{(N)} + f E^{(N+1)} \label{HK_f}
\end{equation}
\begin{equation}
\begin{aligned}
\therefore \mu^{(N)} _+ &= \left.\frac{\partial E[N]}{\partial N}\right|_{N+} = E^{(N+1)}-E^{(N)} = -A^{(N)} \text{   for } 0 < f \leq +1 \\ 
\mu^{(N)}_- &= \left. \frac{\partial E[N]}{\partial N}\right|_{N-} = E^{(N)}-E^{(N-1)} = -I^{(N)} \text{   for } -1 \leq f < 0 \label{IandA}
\end{aligned}
\end{equation}
\begin{equation}
\Delta \mu^{(N)} = I^{(N)}-A^{(N)} = E^{(N+1)} + E^{(N-1)} -2 E^{(N)} \label{del_mu}
\end{equation}

Hence, $\mu^{(N)}$ jumps discontinuously by an amount $\Delta \mu^{(N)}$ whenever crossing the integer electron number $N$. The functional derivative of total energy $\left(\frac{\partial E[N]}{\partial N}\right)$ \tcr{shows similar nature}. Hence, the {\bf derivative discontinuity} occurs at each integer electron number. So, we see that the famous \textbf{\emph{derivative discontinuity is a property of all methods following HK theorem, hence, approximations in Kohn-Sham DFT should satisfy that property.}} From Eq. [\ref{HK_f}], the {\bf piecewise linearity} condition is found as
 $E^{(N+f)} = E^{(N)}- f \{ E^{(N)} - E^{(N+1)} \} $ which further implies that $E^{(N+f)}+E^{(N-f)} > 2E^{(N)} \Rightarrow I^{(N+1)} < I^{(N)}$. Fig.\ref{fig1} may provide more insight.
\end{svgraybox}

\section{The Kohn-Sham (KS) DFT}
HK theorem allows the total energy $E[\rho_0]$ to exactly determine the ground state density $\rho_0$ via the variational Eq. [\ref{eq_HK2}], which is sufficient to determine the ground state properties. But, no direct expression of $E[\rho_0]$ \tcr{can be found using} HK theorem. In KS scheme, local single particle effective potential $v_{eff}(\vr)=v_{ext}(\vr)+v_{ee}(\vr)$ is found which corresponds to the exact ground state density $\rzr$ of the interacting system. This requires an assumption that the $v$-representable ground state density of interacting system $\rzr$ is also $v$-representable in a non-interacting system $S$. Once single particle effective potential is found the coupled equation for $N$ interacting electrons transforms to a single particle equation, called the KS equation. \tcr{The eigenstates found as the solution of KS equation are the KS spin-orbitals. In ground state (GS), the $N$ electrons occupy the lowest lying $N$ spin-orbitals. Remember that, each energy eigenstate is doubly degenerate for up-down spin states. So, from total energy, using KS scheme, we can find a single particle \schr as:} 
\begin{align}
&[-\frac{1}{2}\nabla^2 + v_{eff}(\vr)] \psi_i([\rho_0], \vx) = \epsilon_i \psi_i({\bf x}) 
~~~ \text{where, }~~~ [\epsilon_1 \leqslant \epsilon_2 \leqslant ...] 
\label{eq_KS}\\
& \text{The kinetic energy:} ~
 T^S[\rho_0] =\sum_{i=1}^N \int \psi^\ast_i([\rho_0], \vx) (-\frac{1}{2}\nabla^2 ) \psi_i([\rho_0], \vx) d\vx \label{T}\\
& \text{The GS density:}~ \rzr =  \sum_{i=1}^N \int \psi^\ast_i([\rho_0], \vx) \psi_i([\rho_0], \vx) d\vx
\end{align}

\emph{As, $T^S[\rho]$ is the kinetic energy of $S$ system, so, in the KS e-e interaction energy $E^{KS}_{ee}[\rho_0]$, correlation-kinetic effect also contributes along with the combined effect (already present in xc) of Pauli principle and Coulomb repulsion. However, KS-DFT does not describe their individual contribution in $E^{KS}_{ee}[\rho_0]$, whereas, in Q-DFT framework these terms are exactly defined (see, article 3.4 in \citep{sahni2016}).}  Since, the Coulomb part of $e-e$ interaction energy  is the {\bf Hartree energy} ($E_H$) of Eq. [\ref{E_H}], we can write $E^{KS}_{ee}[\rho_0]$ as:
\begin{align}
	E^{KS}_{ee}[\rho_0] = E_{H}[\rho_0] +E_{xc}^{KS}[\rho_0] ~~\&~~ v_{ee}(\vr) = v_{H} (\vr) +v_{xc} (\vr) \\
	v_{H} ([\rho_0], \vr) = \frac{\partial E_{H}[\rho_0]}{\partial \rzr} 
	~;~~  v_{xc}(\vr)=v_{x} (\vr) + v_{c} (\vr) = \frac{\partial E_{xc}^{KS}[\rho_0]}{\partial \rzr} \label{v_KS_1} \\
	v_{eff}= v_{ext}+v_{ee} = v_{ext} + v_{H} +v_{xc} = v_{ext} + v_{H} + v_{x} + v_{c} \label{v_KS}
\end{align}

Finally, the { \bf Kohn-Sham total energy} using the definitions in Eq. [\ref{T}] and Eq. [\ref{E_H}] :
\begin{align}
E^{KS} = T^S[\rho_0] + \int \rzr v_{ext}(\vr) d \vr + E_H[\rho_0] + E_{xc}^{KS}[\rho_0]  \label{E_KS}
\end{align}	 
This $E_{xc}^{KS}[\rho_0]$ is not explicitly known and requires further approximations as described in Section \ref{sec_Jacob}.

\subsection{Generalised KS Schemes}\label{sec_GKS}
We have already discussed that the $\Psi([\rho_0], \vx)$ can be found from $\rzr$ through constrained search formalism, as, there is no unique $B^{-1}$ mapping available (Please look into Sec. \ref{sec_HK}). This can be done by minimizing the universal density functional $F^{S}$ (of Eq. [\ref{F_HK}]) using Slater type trial wavefunction $\Psi^{S,t}$ that corresponds to $\rzr$. There can be infinite number of such trial wavefunctions. In generalised KS scheme, one searches for the infimum value of the expectation $\langle(\hat{T} +\hat{V}_{ee})\rangle$ with respect to all $N$ dimensional single Slater wavefunctions $\Psi^{S,t}$ that yields the ground state density $\rzr$.
\begin{equation}\label{F^S}
F^S[\rho_0] = \inf_{\Psi^{S,t} \rightarrow \rzr}\langle \Psi^{S,t} | (\hat{T} +\hat{V}_{ee}) | \Psi^{S,t}\rangle
\end{equation}

As, $\hat{V}_{ee}$ is not explicitly known in $S$ system, in standard KS scheme $\langle\hat{T}\rangle$ is used instead of $\langle(\hat{T} +\hat{V}_{ee})\rangle$. Also, this allows to use single Slater wavefunction which is for non-interacting systems. Other two examples of generalised KS (GKS) schemes are Hartree-Fock KS scheme and screened non-local exchange (sx-LDA) scheme, those take a part of $\hat{V}_{ee}$ into account while minimizing \citep{seidl1996}:
\begin{equation}
\begin{aligned}
&\text{KS: }~~~~~~ F^{S}[\rho_0] = \inf_{\Psi^t \rightarrow \rzr}
\langle \Psi^t | \hat{T}| \Psi^t\rangle  \\
&\text{HF KS: }~~~~~~ F^{S}[\rho_0] = \inf_{\Psi^t \rightarrow \rzr}
\left( \langle \Psi^t | \hat{T}| \Psi^t\rangle + E_H[\{\psi_i\}] +E_x[\{\psi_i\}] \right) \\
&\text{sx-LDA: }~~~~ F^{S}[\rho_0] = \inf_{\Psi^t \rightarrow \rzr}
\left( \langle \Psi^t | \hat{T}| \Psi^t\rangle + E_H[\{\psi_i\}] +E^{sx}_x[\{\psi_i\}] \right)  
\end{aligned}
\end{equation}

Here, the exchange energy $E_x[\{\psi_i\}]$ is not the general HF exchange and $E^{sx}_x[\{\psi_i\}]$ is a screened exchange interaction.


\subsection{KS DFT for Fractional $N$; Derivative Discontinuity and Bandgap:}
With the above background, we now move on to discuss the origin of the underestimation of bandgap in KS scheme.
The Euler-Lagrange equations (Eq. [\ref{Euler_N}]) for KS equivalent $S$ system having fractional $N \pm f$ number of electrons are expressed using Eq. [\ref{v_KS_1}-\ref{E_KS}] as \citep{perdew1997}:
\begin{align}
\left. \frac{\delta T^S}{\delta \rho} \right|_{N+f} +v_{ext}+ v_{H}+ v^{N+f}_{xc} &= -A^{(N)} \nonumber \\
(-) \phantom{12} \left. \frac{\delta T^S}{\delta \rho} \right|_{N-f} +v_{ext}+ v_{H}+ v^{N-f}_{xc} &= -I^{(N)} \nonumber \\
\cline{1-2}
\left. \frac{\delta T^S}{\delta \rho} \right|_{N+f} -\left. \frac{\delta T^S}{\delta \rho} \right|_{N-f} + v^{N+f}_{xc} - v^{N-f}_{xc} &= I^{(N)}-A^{(N)} 
\end{align}
It is proved for $N=1$ and argued for $N > 1$ \citep{sagvolden2008} that the KS energy bandgap ($E_g^{KS}$), i.e., the HOMO-LUMO (or, VBM-CBM) gap of the $S$ system arises due to the discontinuity of derivative of kinetic energy \citep{sham1983}, i.e.
\begin{align}
E_g^{KS}=\left. \frac{\delta T^S}{\delta \rho} \right|_{N+f} -\left. \frac{\delta T^S}{\delta \rho} \right|_{N-f} 
&= \epsilon^{(N)}_{N+1} - \epsilon^{(N)}_{N} = \epsilon^{(N)} _{LUMO} -\epsilon^{(N)} _{HOMO}\\
\therefore I^{(N)}-A^{(N)} = \epsilon^{(N)}_{N+1} - \epsilon^{(N)}_{N} &+ v^{N+f}_{xc}  - v^{N-f}_{xc}  \Rightarrow E_g= E_g^{KS} + \Delta v_{xc} \nonumber\\
\&~~~ \Delta v_{xc}
= v^{N+f}_{xc} - v^{N-f}_{xc} &= \left(I^{(N)}-A^{(N)}\right) - \left(\epsilon^{(N)}_{N+1} - \epsilon^{(N)}_{N}\right)
\end{align}

For fractional electron number $(N-f)$, the spin-orbitals $\psi_i(\vx_i)$ of Eq. [\ref{eq_KS}] are occupied completely till $(N-1)$ and the $N$-th spin-orbital should have fractional occupancy of fraction $f$. The ground state energy density $\rzr$, and, the total energy in KS equivalent $S$ system for $(N-f)$ electrons follow the ensemble equation for fractional electron (Eq. [\ref{HK_f}]), so that:
\begin{align}
\rho^{(N-f)} =  \sum_{i}^{N-1} |\psi_i([\rho_0],\vx_i)|^2 + f|\psi_N([\rho_0],\vx_N)|^2 =(1-f) \rho_0^{(N-1)} + f \rho_0^{(N)} \nonumber\\
E^{(N-f)} = (1-f) E^{(N-1)} + f E^{(N)} \Rightarrow \left. \frac{\partial E}{\partial N} \right|_{N-} =  E^{(N)} - E^{(N-1)}
 = \epsilon^{(N)}_N
\end{align}
\tcr{From Eq. [\ref{mu_IA}], we see that} $E^{(N)}-E^{(N-1)}= -I^{(N)}$. So, we finally reach   to the {\bf Ionization Potential (IP) theorem} \citep{perdew1997}:
\begin{itemize}
	\item[] \emph{The highest occupied Kohn-Sham eigenvalue for $N$ electron system (even fractional) is the minus of the ionization energy of the system~; $\epsilon^{(N)}_N = -I^{(N)}$.}
\end{itemize}
 

\section{Jacob's Ladder of Exchange-Correlation Functional} \label{sec_Jacob}

The xc potential $v_{xc}$ or energy $E_{xc}$ \tcr{can not be exactly calculated for many-electron system. There are several levels of approximations used to express the exchange correlation of KS system. It is like climbing a ladder; more rungs one climbs, more accuracy is achieved. Obviously, to achieve that, more effort is needed, in terms of mathematical complexcity, and, computational costing.} From now if spin-orbitals are referred as $\psi(\vx)$ or orbitals as $\phi(\vr)$ they will represent the density dependent versions of KS scheme $\psi([\rho_0],\vx)$ and $\phi([\rho_0], \vr)$, if not explicitly defined otherwise.

\begin{figure}[t!]
	\centering
	\sidecaption
	\includegraphics[scale=0.5]{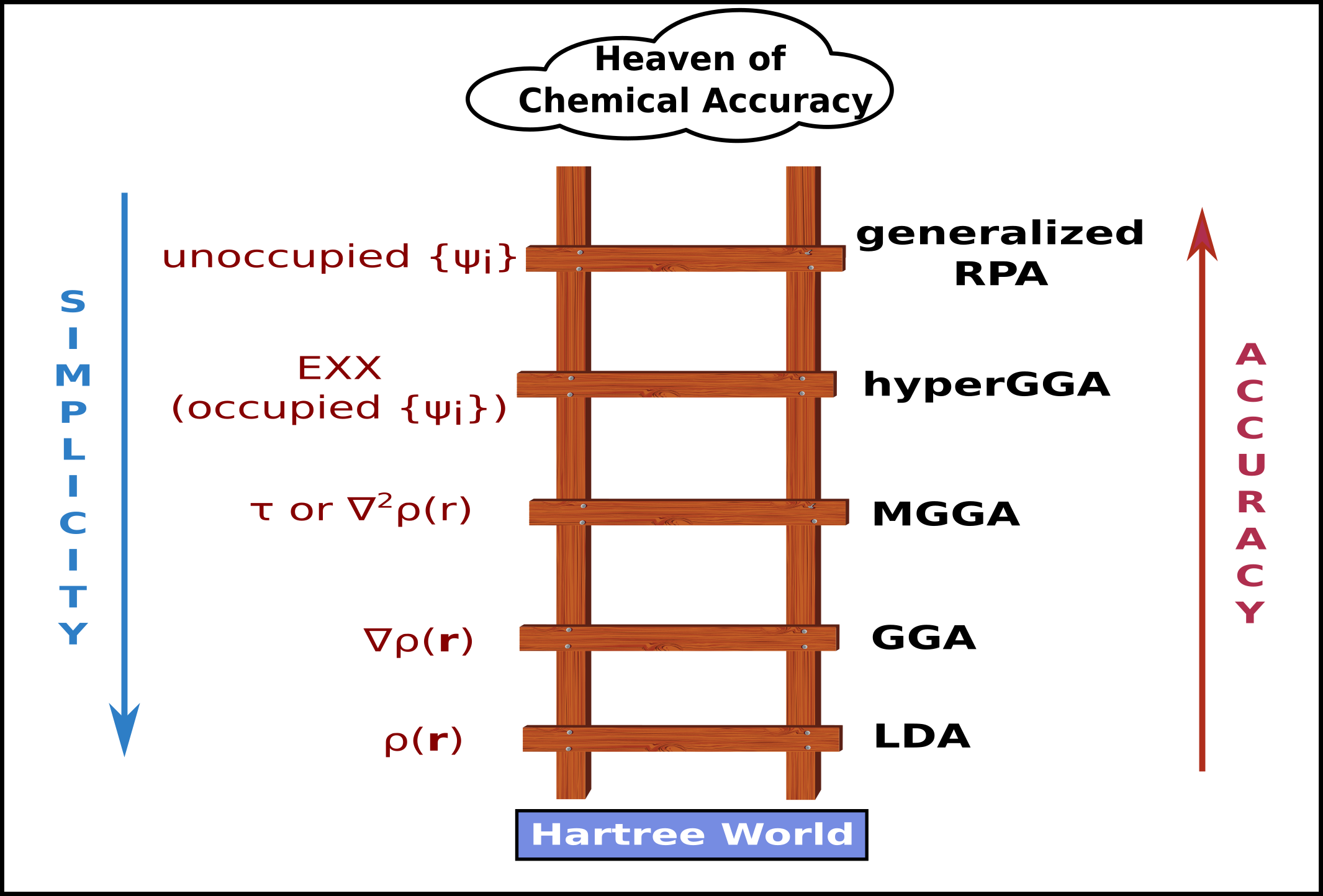}
	\caption{Jacob's Ladder in DFT.}
	\label{fig_Jacobs}       
\end{figure}

The first rung of Jacob's ladder (Fig. \ref{fig_Jacobs}) starts with the introduction of the easiest model for calculating xc, the homogeneous electron gas (HEG) model or Jellium model \citep{perdew2001}. In this model, the non-relativistic electron cloud is taken to be homogenous, i.e., density is everywhere same $\left(\rho_0(\vr)=const.\right)$. A solid state system can be assumed to be a box (Wigner cell) with periodic boundary condition and the electron orbitals can be described by plane waves $\phi_i = \frac{1}{\sqrt{\mathscr{V}_{cell}}}e^{\sqrt{-1}\vk.\vr}$, where $\mathscr{V}_{cell}$ is the volume of the cell. As, $k=\frac{2\pi}{\lambda}$ and electron wavefunction should vanish at cell boundary $\pm L/2$, for one dimension, we can understand that one electron should occupy $2\pi/L$ length in real space. Equivalently, in three dimension it should occupy a volume $\frac{(2\pi)^3}{\mathscr{V}_{cell}}$. Hence, number of electrons in the wave vector range $\vk$ to $\vk+d\vk$, considering both spins, is given by $2\frac{\mathscr{V}_{cell}}{(2\pi)^3}d\vk$ and $d \vk = 4 \pi k^2 dk$. The total number of electron $N$, and, the electron density $\rho$ can be found by integrating $dk$ till the Fermi wave vector $k_F$ corresponding to the highest occupied energy, i.e., Fermi energy $E_F$. 
\begin{equation}
\begin{aligned}
N=\int_{0}^{k_F} 2\frac{\mathscr{V}_{cell}}{(2\pi)^3} 4 \pi k^2 dk = \frac{\mathscr{V}_{cell} k_F^3}{3\pi^2} 
~;~~\rho_0  = \frac{k_F^3}{3\pi^2} \Rightarrow k_F = \{3\pi^2 \rho_0\}^{1/3} 
\propto \rho_0^{1/3}
\end{aligned}
\end{equation}

\subsection{Local Density Approximation (LDA)}
In local density approximation, the electron gas is considered to be locally uniform ($\rzr=\rho_0$) as in HEG, and, the exchange energy is expressed in term of exchange energy per particle $\varepsilon_{x}^{HEG}$ or in term of exchange energy density $\tilde{\varepsilon}_{x}^{HEG}=\varepsilon_{x}^{HEG}\rho_0$ (for derivation of $\varepsilon_{x}^{HEG}$ value, see Sec. 4.3 of Ref. \citep{engel2013}) as:
\begin{eqnarray}
&	E_x^{LDA}[\rho] =\int \varepsilon_x^{HEG}[\rho_0]~ \rho_0  d\vr ~;~ \varepsilon_x^{HEG} [\rho_0] = -\frac{3k_F(\vr)}{4\pi} 
	= -\frac{3}{4}\left( \frac{3}{\pi}\right)^{1/3}\rho_0^{1/3} \\
&\Rightarrow E_x^{LDA}[\rho] = -\frac{3}{4}\left( \frac{3}{\pi}\right)^{1/3} \int \rho_0^{4/3} d\vr 
	~~\& ~~ v_x^{LDA}[\rho_0]= -\left( \frac{3}{\pi}\right)^{1/3} \rho_0^{1/3} = - \frac{k_F}{\pi} \label{xc_LDA}
\end{eqnarray}

If we take a sphere where one electron is present, then the density can be presented as $\rho_0 = \frac{1}{4/3\pi r_w^3}$. Such sphere is called {\bf Wigner-Seitz sphere} of radius $r_w$.
\begin{equation}
	r_w= \left(\frac{3}{4\pi \rho_0}\right)^{1/3} = \left(\frac{9\pi}{4}\right)^{1/3}\frac{1}{k_F}
	 ~~\&~~~ \varepsilon_x^{HEG}[\rho_0] = -\frac{3}{4\pi} \left(\frac{9\pi}{4}\right)^{1/3}\frac{1}{r_w}
\end{equation} 

The correlation is exactly known for two extreme limits, high density weak coupling limit ($r_w\rightarrow 0$) and low density strong coupling limit ($r_w\rightarrow \infty$).
\begin{eqnarray}
	\left. \varepsilon_c \right|_{r_w\rightarrow 0} &=& c_0 \ln r_w  - c_1 + c_2 r_w \ln r_w - c_3 r_w + ... \nonumber\\
	\left. \varepsilon_c \right|_{r_w\rightarrow \infty} &=& -\frac{d_0}{r_w} + \frac{d_1}{r_w^{3/2}} -\frac{d_2}{r_w^2} + ...
\end{eqnarray}

For high density limit, the leading contribution can be calculated using RPA, and, the next higher level contribution is calculated using first order response function (second order exchange). The first two positive constants are $c_0 = 0.031091$  and $c_1 = 0.0467$. A complete RPA calculation is provided by von Barth and Hedin, as well as, by Vosko, Wilk and Nusair \citep{von1972,vosko1980}. For, low density limit, the first two constants $d_0$ and $d_1$ can be estimated from the (Madelung) electrostatic energy and the zero point vibrational energies of the Wigner crystal. For real system having intermediate range of densities, the simplest approach is to interpolate in-between, and, there are several methods developed over years \citep{vosko1980,perdew1981}. A systematic discussion on the development of LDA exchange is presented in Chap-6 of \citep{dreizler2012}. Generalization of LDA for spin polarized system is known as LSDA, and, the exchange is $\varepsilon^{HEG}[\rho_{0\uparrow},\rho_{0\downarrow}]$.
As the HEG model better suits solids than molecules, LSDA performs better in solids. The bond lengths and lattice constants \tcr{predicted by LSDA} are manageable. But in the case of band gap or ionization energy estimation, the scenario is far worse.

To understand the reason, let us go back to the exchange correlation potential in Eq. [\ref{xc_LDA}]. As soon as the density $\rzr$ starts to fall exponentially as $\rzr \sim e^{-\alpha r}$, the exchange potential $v_x^{LDA}[\rho_0] \sim e^{-\alpha r/3}$. In asymptotic region ($r \rightarrow \infty$) the the exponential fall of exchange potential violates the condition $\lim\limits_{r\rightarrow \infty} v_x \sim \frac{1}{r}$ \citep{levy1984}. Similar problem is experienced for correlation which should follow the power law nature \citep{almbladh1985}. Another factor of the failure of LDA is termed as {\bf Delocalization error} which is discussed in [B3].

\subsection{Generalised Gradiant Approximation (GGA)}
LDA uses homogeneous electron gas model, but electron density can never be homogenous in any real system. As a starting point HEG model is good, but then, one have to introduce next level of corrections. The general choice is the slowly varying HEG model for which $\varepsilon_x$ can be calculated using {\bf gradient expansion approximation (GEA)} for high density limit as:
\begin{equation}
E_x^{GEA} = \int \varepsilon_x^{HEG}[\rho_0]~\rho_0 \left[ 1 + \tilde{\mu} s^2 + \cdots \right] d\vr ~; ~~ 
 s = \frac{|\nabla \rho_0|}{2 k_F \rho_0} =\frac{|\nabla \rho_0|}{2\{3\pi^2\rho_0\}^{1/3}\rho_0} \label{E_x_GEA}
\end{equation}

Here $\{\tilde{\mu},\cdots\}$ are constants which can be calculated from GEA. Antoniewicz and Kleinman have calculated $\tilde{\mu}^{GEA}=10/81$ \citep{antoniewicz1985}.
Though, the mathematical derivation is rigorous for GEA, however, for atoms and molecules, the slowly varying idea can  not be justified. But, it has given a clue, and, finally Perdew and Wang (PW91) \citep{PW91} have made the generalised gradient approximation simple by introducing analytical parameters. \tcr{Further improvements to this have lead to the Perdew, Bruke, Ernzerhof (PBE) form \citep{PBE} of GGA. Using {\bf exchange enhancement factor} $\tilde{f}(s)$, $E_x^{GGA}$ becomes:}
\begin{align}
	E_x^{GGA} = \int  \varepsilon_x^{HEG}[\rho_0] ~\rho_0 \tilde{f}'(s) d\vr 
	= -\frac{3}{4}\left( \frac{3}{\pi}\right)^{1/3} \int  \rho_0^{4/3} \tilde{f}(s) d\vr \label{E_x_GGA}\\
	\tilde{f}^{PBE}(s) = 1 + \frac{\tilde{\mu} s^2}{1+\tilde{\mu} s^2/\kappa} ~;~~ \tilde{\mu} =  0.21951 \approx 1.78 \tilde{\mu}^{GEA}, \kappa \leq 0.804 \label{f_PBE}
\end{align}

The exchange enhancement factor, which is generally expressed as a power series of $s$, recovers the uniform gas limit for $\tilde{f}(s)=1$. The $\tilde{\mu} \approx 2 \tilde{\mu}^{GEA}$ is followed by PW91, PBE and other GGA functionals and is appropriate for expressing exchange energies of neutral atoms. But in solids the densities are almost slowly varying, so $\tilde{\mu} = \tilde{\mu}^{GEA}$ is a better choice. Thus, in the modified version of PBE for solids (PBEsol) $\tilde{\mu}$ is taken as $\mu^{GEA}$ \citep{PBEsol}. 

 The density falls exponentially as $\rzr \sim e^{-\alpha r}$ ($\alpha$ is decay constant) and $\alpha$ can be expressed in term of chemical potential $\mu_-^{(N)}$ (\emph{do not get confused with $\tilde{\mu}$ of Eq. [\ref{E_x_GEA}]}) \citep{vLB, levy1984}. Thus, Eq. [\ref{IandA}], $\mu_-^{(N)}=-I^{(N)}$, and, the IP theorem $\epsilon^{(N)}_N = -I^{(N)}$ yield: 
\begin{equation}
\rzr = Ne^{-\alpha r}, \text{ where,  } \alpha = 2 \sqrt{-2\mu_-^{(N)}} =  2 \sqrt{2I^{(N)}} = 2 \sqrt{-2\epsilon^{(N)}_N} \label{rho_eq}
\end{equation}

So, density at asymptotic region depends on the HOMO eigenvalue $\epsilon^{(N)}_N$. Now, $\rho_0$ from  Eq. [\ref{rho_eq}] is inserted into the Eq. [\ref{E_x_GGA}] to finds $E_x^{GGA}$, and, the corresponding potential $v_x^{GGA}=\frac{\partial E_x^{GGA}}{\partial \rzr}$. \tcr{Thus, $\tilde{f}'(s)$ can be found.}

The $v_x^{GGA}$ should follow the asymptotic limit $\lim\limits_{r\rightarrow \infty}v_x \sim -\frac{1}{r}$. This is the starting idea of the semi-empirical GGAs. Furthermore, Savin \etal have showed that for finite systems, the accurate treatment of exchange-correlation potential, particularly in the asymptotic region, leads to eigenvalue differences between the HOMO and LUMO energies close to true bandgap \citep{savin1998}. This is the philosophy of {\bf semi-empirical functional} which has been first introduced by Becke (Becke86) \citep{becke1986}. Parametrization of Becke86 and PW91 potentials have also been proposed for producing better asymptotic behaviour \citep{harbola2002}.

Similar philosophy of producing exact nature of exchange potential at asymptotic region for better electronic structure prediction has been followed by vanLeeuwen and Baerends (vLB) and the corrected LDA  exchange-correlation potential (also known as LB94) has been built as \citep{vLB}:
\begin{equation}
v_{xc}^{vLB}(\vr)= [v_{x}^{LDA}(\vr)+ v_{x}^{vLB}(\vr)]+v_{c}^{LDA}(\vr)  ~;~~
v_{x}^{vLB}(\vr)= - \frac{\beta z^2 \rho^{1/3}(\vr) }{1+3 \beta\ z \sinh^{-1} (z)} 
\label{eq6}
\end{equation}
Here, $z=\frac{|{\nabla} \rho(\vr) |} {\rho ^{4/3}(\vr)} = \frac{s}{2\{3\pi^2\}^{1/3}}$ and $\beta = 0.05$,
It has been employed to study the effect of the correct asymptotic behaviour of potential on the response properties of atoms \citep{banerjee1999}, and, is a very successful potential in describing different properties (e.g, photoionization, absorption spectra, etc.) in atomic and molecular physics due to its correct asymptotic behaviour \citep{stener2000, pertot2017}. The vLB corrected LDA has been first applied to solids by Singh \etal \citep{singh2013, singh2015}. In its original form, the $\beta$ has been taken as constant for every material, which may not be always logical.
In \citep{singh2016}, Singh \etal have varied the $\beta$ using the IP theorem ($\epsilon^{(N)}_N = -I^{(N)}$), and, have applied to diverse set of materials, which indeed becomes a successful method for proper bandgap prediction.
In this IP-vLB work, for optimal choice,  $\beta$ has been varied until the highest-occupied molecular orbital (HOMO) eigenvalue $\epsilon_{N}^{(N)}$ matches with the minus of ionization energy ($-I^{(N)}$).

The material dependency can also be availed through the muffin tin radius in localised orbital schemes, or, in better form in full potential NMTO (FP-NMTO) method through a combination of hard sphere and muffin tin radii \citep{NMTO, NMTO1}.
The minimal basis set used in FP-NMTO method is more accurate and flexible than LMTO-ASA. Being a full-potential method, it can handle the interstitial region more accurately. The larger overlap between muffin tin orbitals in FP-NMTO can produce the curvature of potential in the region in-between atoms can be produced more precisely. Also, the introduction of higher order energy derivatives in the process of removing the energy dependency of the basis set minimizes the error.
This has been utilised by Datta \etal by incorporating the original vLB potential in self-consistent FP-NMTO, which produces almost exact bandgaps for traditional group IV (Si, Ge) and group III-V (GaAs, InP, etc.) semiconductors \citep{datta2019}. This method is as fast as any other GGA calculation. So, this type of potential-only correction which is not coming from any functional derivative of energy can also provide better electronic structure prediction with tricky tuning as in \citep{singh2016, datta2019}.

\begin{figure}[h!]
	\centering
	\sidecaption
	{\includegraphics[scale=0.6]{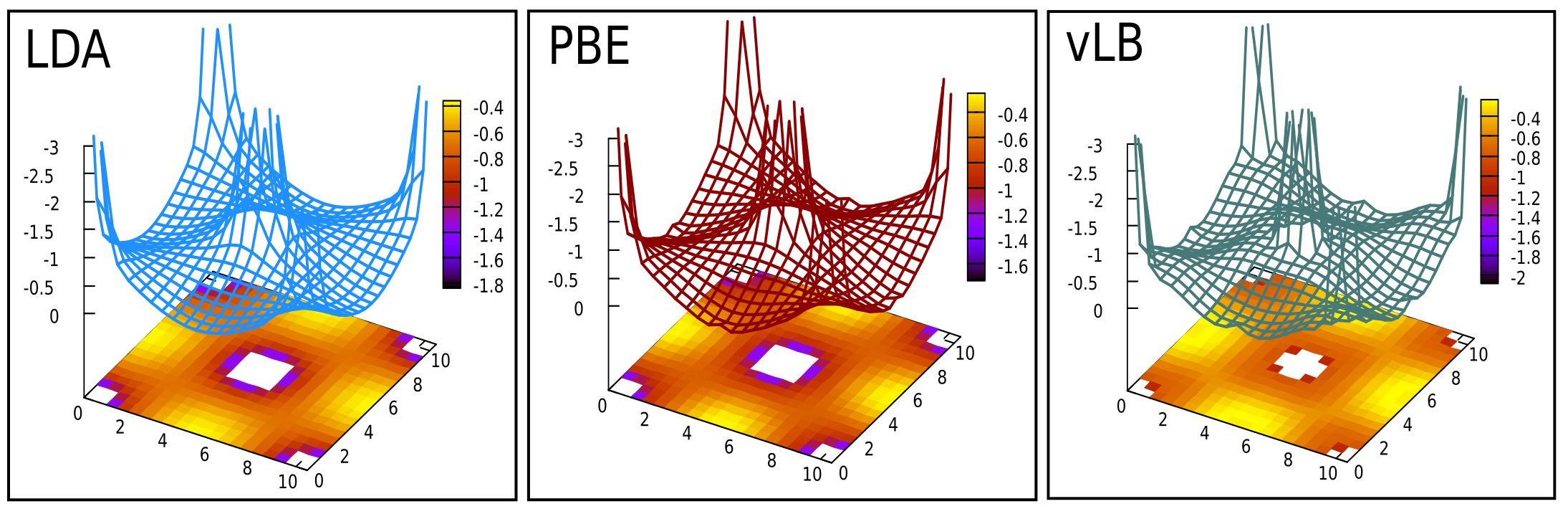}}
	\caption {FP-NMTO XC potential (z-axis is range) for LDA, PBE and LDA+vLB in $\left[1 0 0\right]$-plane.}
	\label{fig_xc-2D}       
\end{figure}

A comparison of LDA, PBE and vLB xc potential in $[100]$ plane for Ge, calculated using FP-NMTO method, is presented in Fig. \ref{fig_xc-2D}  \citep{datta2019}. The LDA potential is smooth, whereas, PBE is showing a variation in the interstitial region (region in-between Ge atoms) and vLB produces larger variation. These variations are is due to the involvement of the gradient correction of charge density.

Direct potential correction has been followed by other methods, as in Becke Johnson (BJ) \citep{becke2006} GGA. Being a derivative of energy functional, the potential can only be defined to within an arbitrary constant, so, $\lim\limits_{r\rightarrow \infty}v_x \sim -\frac{1}{r} + const.$ is also possible. BJ have proposed an effective exchange potential, which is not derived from energy derivative, followed that. They have shifted the exchange potential so that HOMO energy equals to its  exact Hartree-Fock HOMO energy.

Armiento and K{\"u}mmel (AK13) have proposed an exchange enhancement factor $\tilde{f}^{AK13}$, so that, the potential can be found by functional derivative, and then, like the BJ scheme, the asymptotic constant shift of exchange potential has been utilised \citep{armiento2013}.
\begin{equation}
\tilde{f}^{AK13}(s) = 1 + B_1 s  \ln(1+s) + B_2 s \ln(1 + \ln(1+s)) ;~ v_x^{mod} (\vr)= v_x(\vr) - \lim\limits_{r\rightarrow \infty} v_x(\vr)
\end{equation}
Where $B_1 = 3/5\tilde{\mu}^{GEA} + 8\pi/15$ and $B_2 =\tilde{\mu}^{GEA}-B_1$. In this process, the HOMO eigenvalue is also shifted as 
$\alpha = 2 \sqrt{-2\left(\epsilon^{(N)}_N - \lim\limits_{r\rightarrow \infty} v_x(\vr)\right)}$ in Eq. [\ref{rho_eq}].

Later, they have pointed out some drawbacks of such nonvanishing asymptotic exchange potentials, those demand constant shift, though, the proper production of derivative discontinuity made these successful in reproducing atomic and molecular properties, and, atomic-shell structures. Bandgaps predicted by these are close to experimental values establishing their usefulness in semiconductor theory \citep{aschebrock2017}.

\begin{figure}[h!]
	\centering
	\sidecaption
	\framebox{\includegraphics[scale=0.35]{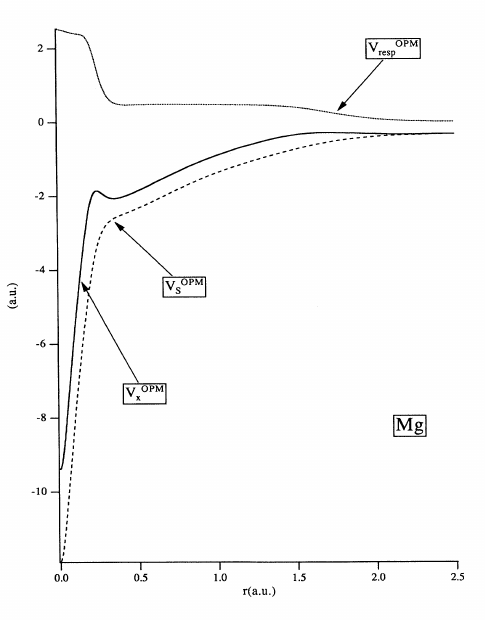}}
	\caption{Variation of exchange potential $v_x(\vr)$ and its components: Slater type exchange potential $v_x^S(\vr)$, its response $v_x^{resp}(\vr)$ for Mg calculated using optimised potential method (see, Sec. \ref{sec_OPM}). The kink in $v_x(\vr)$ produces the non monotonous variation with $\vr$ which is represents the shell structure. Reproduced with permission from \citep{gritsenko1995}}
	\label{fig_GLLB}       
\end{figure}

Gritsenko \etal (GLLB) have noticed from the optimised potential method (OPM) result that, the shell closure like nature of exchange potential can be reproduced by a combination of two terms; one is a screening type Slater exchange  (Eq. [\ref{g_x}]]) originating from Fermi exchange hole, and, another is the response of exchange only pair-correlation function $g_x(\vr,\vr')$ of Eq. [\ref{g_x}], written as \citep{gritsenko1995}:
\begin{align}
v_x (\vr) &= v_x^{S}(\vr) + v_x^{resp}(\vr) \\
\text{where,}~~ v_x^{resp}(\vr)& = \frac{1}{2} \int \rho_0(\vr')  \left[ \int \frac{\rho_0(\vr'')}{|\vr'-\vr''|} 
 \frac{\delta g_x(\vr',\vr'') }{\delta \rzr} d\vr'' \right] d\vr' 
\end{align} 

GLLB have approximated $v_x^{resp}(\vr)$ using HOMO energy  $\epsilon^{(N)}_N$ and eigenstates $\{\psi_i\}$ as:
\begin{equation}
v_x^{resp, GLLB}(\vr) \simeq K_{g} \sum_{i}^N \sqrt{\epsilon^{(N)}_N - \epsilon^{(N)}_i} \frac{|\psi_i(\vx)|^2}{\rzr}
\end{equation}
where, $K_{g}$ is calculated from HEG model as $8\sqrt{2}/3\pi^2$. Kuisma \etal have combined PBEsol correlation with this GLLB exchange to produce GLLB-SC $xc$ potential aimed for solids \citep{kuisma2010}:
\begin{align}
v_{xc}^{GLLB-SC}(\vr) = 2\varepsilon_x^{PBEsol} (\vr) +
  K_{g} \sum_{i}^N \sqrt{\epsilon^{(N)}_N - \epsilon^{(N)}_i} \frac{|\psi_i(\vx)|^2}{\rzr}
 + v_c^{PBEsol} (\vr)
\end{align}

These GLLB and GLLB-SC potentials produce the derivative discontinuity for solids, because, a small $\delta \rho$ addition to density corresponding integer electron opens up a new orbital with fractional occupation and $\epsilon^{(N)}_N$ jumps. GLLB-SC performs better than GLLB due to incorporation of PBEsol GGA  correlation \citep{tran2018}.

\begin{figure}[]
	\centering
	\sidecaption
	\framebox{\includegraphics[scale=0.32]{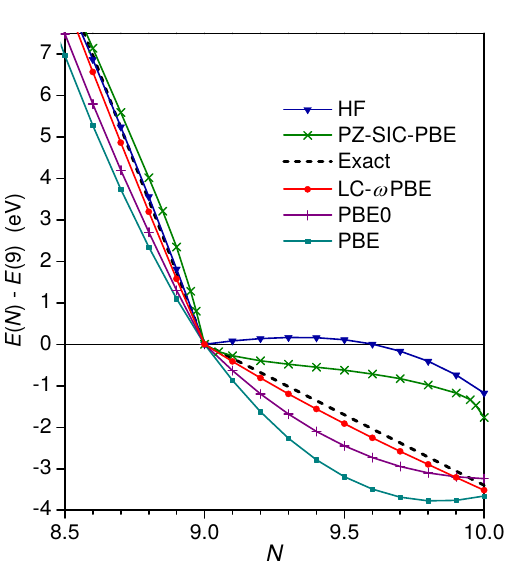}}
	\framebox{\includegraphics[scale=0.41]{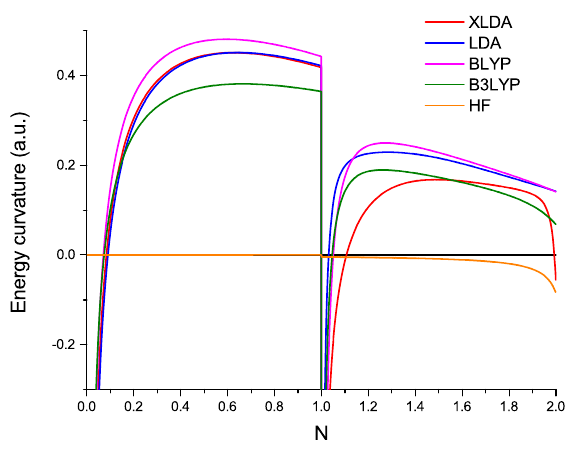}}
	\caption{(left) Total energy ($E$) in eV for F atom as function of electron number $N$ from \citep{vydrov2007}.
		(right) Energy curvature $\frac{\partial^2 E}{\partial \rho^2}$ as a function of $N$ of H atom from \citep{li2017} (XLDA is exchange-only LDA). Reproduced with permission from \citep{vydrov2007,li2017}.}
		\label{fig2}       
\end{figure}

\begin{svgraybox}
\textbf{\large B3: Delocalization Error}

Reproduction of piecewise linearity of total energy of HK systems (see [B2]) is a necessary condition to follow by every (semi-)local approximations. For system with fractional charge, LDA energy ($E$) curve with electron number ($N$) gives a piecewise convex nature. It means these functionals {should calculate very} low energy for fractional charge. As fractional charge arises due to electron exchange between atoms, it signifies the electron delocalization, and, functionals with convex nature wrongly delocalizes electron. So, this error is termed as {\bf Delocalization error} \citep{mori2008}.
This piecewise convex behaviour is present in many semi-local functionals. On the other hand, the HF energy curve is piecewise concave, and, imposing localization of xc kernel through self-interaction correction (SIC), e.g., Perdew-Zunger PZ-SIC, follows similar wrong concave nature (see, left panel of Fig. \ref{fig2}) \citep{mori2008,vydrov2007}. Long range corrected hybrid functional LC-$\omega$PBE produces almost exact piecewise linear character. Besides these numerical evidences, recently the concavity of HF energy is proved mathematically \citep{li2017}. However, the piecewise convexity of semi-local functionals is not guaranteed for all fractions, especially for low fractions, as seen in the right panel of Fig. \ref{fig2}.
\end{svgraybox}

\subsection{Meta-GGA}
LDA starts with the homogeneous distribution of density and GGA introduces the \tcr{effect of} first order spatial variation of density ($\nabla\rho_0$) \tcr{within xc potential}. The next level of correction should naturally come through the introduction of the second order term $\nabla^2 \rho_0$. Some functions have been proposed, but, there is a problem with such terms; the corresponding potential becomes fourth order gradient of density, which blows out near the nuclei making the calculation highly unstable. The kinetic energy density term $\tau(\vr)$  provides an alternative pathway.
\begin{equation}
\tau(\vr)=\sum_\sigma\tau_\sigma(\vr)=\frac{1}{2}\sum_{\sigma}\sum_{i}^{occ}|\nabla \psi_i(\vr,\sigma)|^2
\end{equation}

The gradient expansion of kinetic energy provides a flexibility, we can add  $\nabla^2 \rho_0$ dependent term in kinetic energy density not changing the kinetic energy itself. This is because $\int \nabla^2 \rho_0 d\vr$ vanishes (Gauss' theorem) if integrated over any finite system. Using gradient expansion (may refer to Sec. 4.4.3 of \citep{engel2013}) of kinetic energy density upto second order, one can reach to the expansion \citep{brack1976}:
\begin{equation}
\tau^{GEA}(\vr) = \tau^{HEG}+ \frac{1}{72}\frac{|\nabla \rzr|^2}{\rzr} + \frac{1}{6}\nabla^2 \rzr 
~~;~~ \tau^{HEG} = \frac{3(6\pi^2)^{2/3}}{10} \rho_0 ^{5/3}(\vr) \label{tau_GEA}
\end{equation}
So, kinetic energy density $\tau(\vr)$ can serve as a substitute of $\nabla^2 \rho_0(\vr)$, and, the story of meta-GGA (MGGA) begins. In MGGA, the exchange enhancement factor is defined as: $\tilde{f}_x^{MGGA} \equiv \tilde{f}_x(\rho_0, \nabla\rho_0, \tau)$.

Perdew, Kurth, Zupan and Blaha (PKZB) have further used the idea of PBE GGA to develop a semi-empirical meta-GGA satisfying spin-scaling, uniform density scaling and Lieb-Oxford lower bound constraints \citep{PZKB}. \tcr{A list of constraints used in DFT approximations can be found in Ref. \citep{SCAN}.} In reduced functional of $p$ and $\tilde{q}$, $\tilde{f}_x$ is expressed as:
\begin{equation}
\begin{aligned}
\tilde{f}^{PKZB}_x(p,\tilde{q}) = 1+ \frac{\zeta}{1+\zeta/\kappa} ~~\text{where, }~~ p=s^2 ,~ \tilde{q} &= \frac{3\tau}{2(3\pi^2)^{2/3} \rho_0^{5/3}} - \frac{9}{20} - \frac{p}{12} \label{f_MGGA}\\
\zeta^{PZKB} = \frac{10}{81}p + \frac{146}{2025} \tilde{q}^2 - \frac{73}{405} p \tilde{q} &+ \left[D+\frac{1}{\kappa}\left(\frac{10}{81}\right)^2\right]p^2
\end{aligned}
\end{equation}

The first two term in $\zeta^{PZKB}$ do not contain $\kappa$ , so that the exchange enhancement factor recovers the slowly varying limit of gradient expansion (GEA) of enhancement factor up to the fourth order of $\nabla \rzr$ as:
\begin{equation}
\tilde{f}_x^{GEA} = 1+\frac{10}{81}p + \frac{146}{2025} \tilde{q}^2 - \frac{73}{405} p \tilde{q} + Dp^2 + \cdots \label{F_x_GEA}
\end{equation}

In PKZB scheme, the empirical parameter $D=0.113$ is estimated by minimizing the mean absolute error of the atomization energies for a set of molecules. PKZB functional successfully determines the surface and atomization energies but it overestimates the bond lengths. Also, it remains unsuccessful in describing hydrogen bonded systems.

Tao \etal have attributed this failure in bond length prediction to the PKZB exchange, and, have proposed a corrected exchange called TPSS (Tao-Perdew-Staroverov-Scuseria) \citep{TPSS}. TPSS is the first non-empirical MGGA which makes the empirical parameter of PKZB, $D=0$. Further it includes an additional constraint: making exchange potential finite at the nucleus for the ground state of one and two electronic system. The single orbital kinetic energy density reduces to the \textbf{Weizs\"{a}cker functional} (for spin polarised system, single orbital means two spin channels) $\tau^W$ and two dimensionless quantities are useful (using $\tau^{HEG}$ of Eq. [\ref{tau_GEA}]):
\begin{equation} \label{eq_SCAN}
\tilde{z} = \frac{\tau^W}{\tau} ~;~~ \tilde{\alpha} = \frac{\tau - \tau^W}{\tau^{HEG}} ~;~~ \text{where, } \tau^W = \frac{|\nabla \rzr|^2}{8 \rzr}
\end{equation}

The {\bf iso-orbital indicator} $\tilde{\alpha}$ differentiate between different local bonding environments: $\tilde{\alpha} = 0$ or $\tilde{z}=1$ represents the covalent single bond, $\tilde{\alpha} \approx 1$ i.e., $\tau - \tau^W \approx \tau^{HEG}$ indicates that after bonding there are enough electron cloud available to replicate HEG, so, represents metallic bonding, and, $\tilde{\alpha} \gg 1$ represents weak bonding \citep{sun2013}. For two electron system ($\tilde{\alpha} = 0$), the PKZB-MGGA enhancement factor of Eq. [\ref{f_MGGA}] reduces to GGA form, like Eq. [\ref{f_PBE}]. The corresponding exchange potential, which has a $\nabla^2 \rzr$ term, diverges at nucleus. To avoid that, the constraint $\left. \frac{d \tilde{f}_x}{d s}\right|_{s=0.376} =0$ has been imposed in TPSS scheme, where, $s=0.376$ is the value of $s$ at nucleus for two-electronic system. Satisfying all these constraints, they have reconstructed the $\zeta$ of Eq. [\ref{f_MGGA}] as a complicated analytical functional: $\zeta[p,\tilde{q}, \tilde{z}]$ \citep{TPSS}.

TPSS functional has improved the \tcr{efficiency of the} estimation of lattice constants for solids as well as the bond lengths for molecules, hydrogen bonded complexes whereas, the mean error in calculation of bulk-moduli of solids increases with respect to PKZB functional.
It is less accurate than the PBE GGA for calculation of critical pressure of structural phase transition of solids as well.

To solve such problems, Sun \etal have introduced a functional which satisfies a larger set of constraints and appropriate norms. They have termed it as strongly constrained and appropriately normed (SCAN) functional \citep{SCAN}. In construction of SCAN functional, they have interpolated $\tilde{f}^{PKZB}$ in-between covalent bonding case ($\tilde{\alpha} = 0$), and, metallic bonding case ($\tilde{\alpha} =1$). \tcr{Then, they have further extrapolated} for hydrogen bonded systems ($\tilde{\alpha} \rightarrow \infty$). There is a switching function defined, which switches between $\tilde{\alpha} = 0$ and $\tilde{\alpha} = 1$ in the expression of $\tilde{f}^{SCAN}$. SCAN functional is successful in describing molecular energies, barrier heights of chemical reactions, as well as, in predicting lattice constants, mechanical stability of solids \citep{sun2016,zhang2018}. However, SCAN functional significantly overestimates the magnetization of elemental ferromagnetic materials (Fe, Co, and Ni). In a recent study, the reason is attributed to the insensitivity of the switching function to $\tilde{\alpha}$ for some particular range, and, oversensitivity in another range \citep{mejia2019}. They have proposed a deorbatalized version of SCAN, called as SCAN-L where $\tilde{\alpha}$ is a function of $[\rho, \nabla \rho, \nabla^2 \rho]$.


We have seen that (Eq. [\ref{ex_density1}] using Eq. [\ref{density_spin}] \& \ref{density}) the Fermi exchange hole $\rho_x(\vr,\vr')$ can be expressed in term of first order density matrix. The Fermi hole is highly delocalized. A density matrix expansion (DME) under general coordinate transformation can make the Fermi hole localized, which reduces the difficulty in modelling the conventional Fermi hole. Tao and Mo (TM) have followed similar type of coordinate transformation as done in \citep{van1998}, and, formulate a MGGA exchange enhancement factor $\tilde{f}_x^{TM-DME}$ \citep{TAO-MO}. In the HEG limit, the exchange energy functional and the Fermi hole from the DME are exact, however, this fails for slowly varying densities. To solve this, they have interpolated between the DME and slowly varying densities, as well as, between the corresponding enhancement factors. The slowly varying enhancement factor $\tilde{f}_x^{TM-SC}$ is found in accordance to the fourth order gradient expansion of $\tilde{f}_x^{GEA}$ (Eq. [\ref{F_x_GEA}]). The TM interpolated enhancement factor using the $\tilde{z}$ dependent weight factor $w$ as:
\begin{align}
\tilde{f}_x^{TM} = w ~\tilde{f}_x^{TM-DME} + (1-w) \tilde{f}_x^{TM-SC} ~;~~ w=\frac{ \tilde{z}^2 + 3 \tilde{z}^3}{(1 + \tilde{z}^3)^2 }
\end{align}

So, TM have used $\tilde{z}$ as the iso-orbital indicator,which can differentiate between single-orbital ($\tilde{z} = 0$) and slowly varying $\tilde{z} \approx 1$ bondings. Another widely used indicator in MGGA is $\tilde{t}^{-1} = \frac{\tau}{\tau^{HEG}}$, which can differentiate between covalent and non-covalent bonding, but, can not identify single-orbital regions. As mentioned earlier, TPSS and SCAN have used $\tilde{\alpha}$, which is better than these two and can be directly related with {\bf electron localization function}: $ELF =\frac{1}{1+\tilde{\alpha}^2}$.

However, $\tilde{\alpha}$ dependent MGGA (e.g., SCAN) faces numerical instability originating  from  sharp oscillations in the xc potential, and, can only be eliminated with very fine grids \citep{furness2019,bartok2019}. To solve the problem, Furness and Sun have proposed a indicator $\tilde{\beta} = \frac{\tau - \tau^W}{\tau+\tau^{HEG}}$ \citep{furness2019}. Bart{\'o}k and Yates have indicated regularised SCAN (rSCAN) functional, \tcr{where the switching function is expressed as} $\tilde{\alpha'} = \frac{\tilde{\alpha}^3}{\tilde{\alpha}^2 + 0.001}$. They have used $\tau^{HEG}+10^{-4}$ instead of $\tau^{HEG}$ in Eq. [\ref{eq_SCAN}] to regularize the $\tilde{\alpha}$ for only very small values \citep{bartok2019}. rSCAN improves numerical performance of SCAN at the expense of breaking constraints known from the exact xc functional. In SCAN, LSDA limit for HEG can be recovered through the constraint $\tilde{\alpha} \rightarrow 1$, but, rSCAN loses the correct HEG description. The correct uniform- and non-uniform- scaling properties of $\alpha$, as well as, the correct HEG limit of $E_{xc}$ are restored in r$^2$SCAN \citep{furness2020} using a regularization parameter $\eta$ as:
\begin{equation}
\tilde{\alpha}'' = \frac{\tau - \tau^W}{\tau^{HEG} + \eta \tau^W} ~;~~ \text{where, } \eta = 10^{-3}
\end{equation}

In MGGA regime, potential-only functionals (those are not functional derivative of exchange energy, like vLB in GGA section) are available as well. The Becke Johnson (BJ) semi-empirical potential is given as \citep{becke2006}: 
\begin{equation}
v_{x,\sigma}^{BJ} = v_{x}^{BR}(\vr) + \frac{1}{\pi}\sqrt{\frac{5}{6}}\frac{\tau(\vr)}{\rzr}
\end{equation}

Here $v_{x}^{BR}$ is Becke Russel GGA potential (see, \citep{becke1989}). As the BJ potential have not been formed in the conventional way as any other MGGA functional, so, some veterans in this field do not want to consider it as MGGA \cite{tran2019}. Tran and Blaha have modified the BJ potential (TBmBJ) by introducing a density gradient dependent weight factor $w$ as \citep{tran2009}:
\begin{align}
v_{x,\sigma}^{TBmBJ} = w~ v_{x}^{BR}(\vr) +& (3w-2) \frac{1}{\pi}\sqrt{\frac{5}{6}}\frac{\tau(\vr)}{\rzr} \\
\text{where,}~~ w= -0.012 + 1.023 \sqrt{\mathscr{G}} 
~~&;~~ \mathscr{G}=\frac{1}{{\mathscr{V}}_{cell}} \int_{cell} \frac{1}{2} \frac{|\nabla\rho_0(\vr')|}{\rho_0(\vr')} d\vr' \nonumber
\end{align}
Here, the constant $1.023$ is in (Bohr$^{1/2}$) unit. TBmBJ functional shows great improvement in bandgap prediction, and, optical properties calculations for semiconductors and insulators which will be discussed in the next section.

\subsection{Hybrid Functional Method}
The LDA, GGA , MGGA are the so called (semi-)local functionals, as the exchange energy density at any point $\vr$ depends on the density, its first and second order gradients, and, atmost on the orbitals and/or their gradient, but, only at that point $\vr$. Whereas, for any real system, the exchange energy is expressed in term of density as in Eq. [\ref{E_x_HF}]. Due to the non-local nature of the exchange hole, the exchange energy integrand is fully non-local. This is why the fourth rung of the ladder is important, which includes the \emph{exact exchange} of HF or similar orbital dependent schemes. Another benefit of HF scheme is that it is always free form self interaction error, which is already discussed in Sec. \ref{sec_HF}.
The idea of mixing one part of HF exchange with LDA/GGA/MGGA density functional approximations (DFA) has come from the adiabatic connection formulation (see Sec. 4.4.2 of \citep{engel2013}) and the exchange-correlation energy in hybrid functional scheme is expressed as:
\begin{equation}
E_{xc}^{hybrid} = E_{xc}^{DFA} + w~ \{E_{x}^{HF} - E_{x}^{DFA}\}
\end{equation}
In {\bf global hybrids}, weight factor $w$ is taken as constant, so, applicable over all space. Some popular global hybrids are B3LYP \citep{B3LYP}, non-empirical PBE0 \citep{PBE0} or TPSSh MGGA \citep{TPSSh} hybrids. While global hybrids bring improvement over semi-local DFAs, the inclusion of exact HF type exchange in extended systems is computationally problematic.

To reduce the computational complexity and to achieve better results, the weight factor $w$ \tcr{has been taken as} a function of $\vr - \vr'$ by separating in short range (SR) and long range (LR) pieces in {\bf range separated hybrids} (RSH) as:
\begin{equation}
E_{xc}^{RSH} = E_{xc}^{DFA} + w_{SR} \{E_{x}^{SR-HF} - E_{x}^{SR-DFA}\} + w_{LR} \{E_{x}^{LR-HF} - E_{x}^{LR-DFA}\}
\end{equation}

A popular choice of range separation for the electron-electron repulsion operator is by using general error function $\erf(\vartheta~|\vr-\vr'| )$ as:
\begin{equation}
\frac{1}{|\vr-\vr'|} = \underbrace{\frac{1-\erf(\vartheta~|\vr-\vr'| )}{|\vr-\vr'|}}_{SR} + 
\underbrace{\frac{\erf(\vartheta~|\vr-\vr'| }{|\vr-\vr'|}}_{LR}
\end{equation}
Here, the parameter $\vartheta$ controls the range of separation, as, $\vartheta \rightarrow (0/\infty) \equiv(LR/SR)$. In {\bf screened hybrids} the LR part is ignored, and, the hybridization with HF like functions is done in short range only. This reduces the computational cost significantly. Heyd, Scuseria and Ernzerhof (HSE) have introduced screened hybrid, which uses the error function based range separation in short range only, and, $w_{SR}^{HSE} = 0.25$ \citep{HSE}. In the modified HSE06 hybrid $\vartheta=0.11 ~a_0^{-1}$, where, $a_0$ is the Bohr radius \citep{HSE1,HSE06}.

Short range hybrids are computationally less demanding than the global counterpart for larger systems, but, the short range correction may not provide exact asymptotic nature for molecules. Quantities sensitive to the long-range exchange potential, as well as, to the self-interaction error in density tails may not be addressed well in SR hybrids. Some long range hybrids have been proposed (e.g., LC-$\omega$PBE \citep{vydrov2007}) and
Henderson, Izmaylov, Scuseria, and Savin (HISS) have proposed a three range separated hybrid, where, a mid range variant is added \citep{HISS}.

There is yet another genre of hybrids, the {\bf local hybrids} (LH), where, instead of exchange energies $E_x$, exchange energy densities ($\tilde{\varepsilon}_x$) are mixed locally in space:
\begin{equation}
E_{xc}^{LH} = E_{xc}^{DFA} + \int w(\vr) ~\{\tilde{\varepsilon}_{x}^{HF}(\vr) - \tilde{\varepsilon}_{x}^{DFA} (\vr)\} d\vr
\end{equation}

The problems with local hybrids, and, the recent developments towards the solution is reviewed in Ref. \citep{maier2019}, whereas, the different types of hybrids are compared in \citep{henderson2008,arbuznikov2007}. Performance of global and local hybrids in reduction of many electron self-interaction error is compared, and, it has been found that in smaller systems local hybrids perform better. But, with increasing system size, the performance of local hybrid becomes similar to a global hybrids \citep{schmidt2016}.

\subsection {Random Phase Approximation (RPA) within DFT}
\tcr{The random phase approximation scheme comes the fifth rung of the Jacob's ladder.} RPA is a widely used method in different branches of many-body scattering theory. Within DFT scheme, RPA can be formulated using the adiabatic coupling fluctuation-dissipation theorem (ACFDT), though, this is not the only method. RPA provides the long range correlation exactly, so, it is supposed to provide a good description of electronic structures.  For molecular systems having complex interaction behaviour, this may come to be handy, but for solids, the necessity of RPA is limited. However, the computational cost for RPA calculation is still so high that for large systems it is not widely used yet. For more mathematical details on ACFDT-RPA method, readers are referred to \citep{ren2012}.

\section{Orbital Based Exchange Correlation} \label{sec_OPM}
In KS-DFT the ground state energy $E[\rho_0]$ is a functional of $\rzr$, and, the effective potential of $N$ electronic  $S$ system $v_{eff}(\vr)$ (Eq. [\ref{v_KS}]) is found by variational minimization of $E[\rho]$ (Eq. [\ref{eq_HK2}]) with respect to variation of $\rho(\vr)$. Since, in $S$ system, spin-orbitals $\{\psi_i(\vx)\}$ are functional of $\rho(\vr)$, so, $E$ can also be expressed as a functional of spin-orbitals as $E[\{\psi_i\}]$, and, energy minimization can also be achieved through the variation of $E[\{\psi_i\}]$ with respect to $v_{eff}(\vr)$. This is the basis of {\bf Optimised Potential Method (OPM)} (see, Chap-2 of \citep{fiolhais2003}). In OPM, an integral equation which determines $v_{eff}$ (including effect of $v_{xc}$) is solved simultaneously with the KS equation. Once again, let, the spin-orbitals be written as combination of space and spin parts: $\psi_i(\vx)=\phi_i(\vr)\chi_1(\sigma)$.

Because the expression of exchange energy $E_x$ for $S$ system is not explicitly known in term of the density, the exchange potential $v_x(\vr)$ can be expressed in term of the orbitals as:
\begin{equation}
v_x([\rho], \vr) = \frac{\delta E_x[\{\phi_i\}]}{\delta \rho(\vr)} =\sum_{i}^{occ}\iint  
\left[\frac{\delta E_x[\{\phi_i\}]}{\delta \phi_i(\vr'')}
 \frac{\delta \phi_i(\vr'')}{\delta v_{eff} (\vr')}+c.c.\right] 
\frac{\delta v_{eff} (\vr')}{\delta \rho(\vr)}d\vr'' d\vr'
\end{equation}

Now, $\frac{1}{\phi_i^\ast(\vr)}\frac{\delta E_x[\{\phi_i\}]}{\delta \phi_i(\vr)} = v_{x,i}(\vr)$ as in HF scheme (Eq. [\ref{XP}]) and in OPM it is explicitly known. Whereas, 
$ \frac{\delta \phi_i(\vr')}{\delta v_{eff} (\vr)} = -G_i(\vr',\vr)\phi_i(\vr)$ is calculated using first-order perturbation theory, where, 
$G_i(\vr,\vr') = \sum\limits_{j,i\ne j} \frac{\phi_j(\vr) \phi_j^\ast(\vr')}{\epsilon_j- \epsilon_i}$ is the Green's function, $\{\epsilon_i\}$ are orbital energies. 
Inverse of $\frac{\delta v_{eff} (\vr')}{\delta \rho(\vr)}$ is the static response function:
 $$\varUpsilon(\vr, \vr')= -\sum\limits_{i}^{occ} \phi_i^\ast (\vr) G_i(\vr,\vr') \phi_i (\vr') +c.c.$$
 
As mentioned, in OPM the total energy is minimised by varying the effective potential, so that, the minimization condition becomes: $\frac{\delta E[\{\psi_i\}]}{\delta v_{eff}(\vr)} =0$. As, $v_{ext}(\vr)$ and $v_H(\vr)$ are exactly known, the minimization condition leads to the integral equation:
\begin{equation}
\sum_i \int [v_x^{OPM}(\vr')-v_{x,i}(\vr')] \phi_i^\ast (\vr') G_i(\vr',\vr) \phi_i(\vr) d\vr' +c.c. = 0
\end{equation}
This integral equation is to be solved to get the minimised energy in OPM. 

Kotani has first carried out systematic study on exact exchange potentials on a variety of semiconductors using OPM method and these studies have showed excellent bandgap matching with experiments \citep{kotani1995,kotani1996}.

Another orbital dependent approach comes from the quantal DFT approach \citep{sahni2016}. In Harbola-Sahni (HS) potential method, the exchange potential is expressed as a line integral over an electric field ${\bf \mathcal{E}}(\vr)$ \citep{harbola1989, sahni1990}:
\begin{equation}
W^{HS}(\vr) = -\int_{\infty}^{\vr} \mathcal{E}(\vr').{\bf dl} ~;~ \text{where,}~~ 
\mathcal{E} = \int \frac{\rho_x(\vr,\vr')}{|\vr-\vr'|^3}(\vr-\vr') d\vr'
\end{equation}
For spherically symmetric densities, as used in atomic sphere approximation (ASA) in LMTO-ASA, this $\mathcal{E}$ is curl free. For nonspherical charge densities, the solenoidal part of $\mathcal{E}$ is related to the difference in kinetic energies between the HF and the HS approaches \citep{sahni1997} and the contribution is numerically insignificant \citep{slamet1994}. The HS potential can directly be derived from \schr \citep{holas1997}. Using virial theorem, the exchange energy is found as:
\begin{equation}
E_x^{HS} = \int \rho (\vr) ~ \vr .\nabla W^{HS} (\vr) d\vr
\end{equation}

\section{Performance Comparison of different DFT Methods}

DFT methods are meant to reduce the computational costing, yet producing exact physical properties of materials \citep{giustino2014}. We have seen that, it is a hard task to balance between the atomic-molecular system and solid state systems, as, the nature of the variation of the charge densities are fundamentally different. 
For molecular systems, a rigorous analysis on the performance of different methods can be found in many texts, e.g, \citep{mardirossian2017}. Here we focus on the solids, specifically, semiconductors.

To compare the performance of different functionals, we are relying on two parameters, mean percentage error MPE=$\frac{100\%}{N}\sum\limits_i^N\frac{x_i-x_i^{expt}}{x_i^{expt}}$, and, mean absolute percentage error MAPE=$\frac{100\%}{N}\sum\limits_i^N\frac{|x_i-x_i^{expt}|}{|x_i^{expt}|}$, where, $\{x_i\}$ are DFT calculated values. In case of more than one experimental values, the average of those are taken as $x_i^{expt}$. Having large MPE or MAPE means large deviation from experimental values, indicating poor performance. Now, there is another matter of concern, the consistency of results. If the absolute value of MPE differs much from MAPE, then it indicates that the result from this particular functional sometimes give larger than experimental values, and, sometimes smaller than that, thus, reducing the consistency. As an example, we can take the case of PBE GGA; it overestimates the lattice constants, and, largely underestimates the bandgap of semiconductors, although, the results are consistent in the sense that, one almost never find a PBE lattice constant lower than the experimental value and a PBE bandgap larger than the experimental value.

Another important thing to mention here is that the implementation of same functional in different DFT packages may produce different results, and in some cases, the implementation is tricky, and, sometimes become package dependent as well. A recent article is on the reproducibility of DFT results has aimed to target this issue, and, worth reading \citep{lejaeghere2016}, although, discussion on every single package can not be included in a paper anyhow. So, we try to mention the method and/or package in the tables wherever possible.

\subsection{Structural Parameters}
\subsubsection{Lattice Constant}
\begin{figure}[h!]
	\centering
	\sidecaption
	\framebox{\includegraphics[scale=0.35]{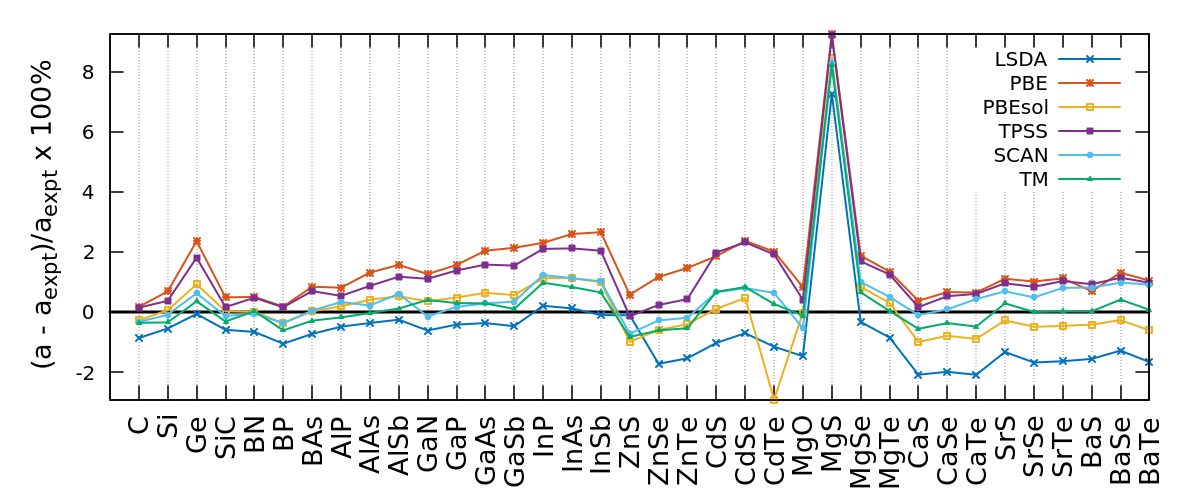}}
	\caption{Relative error percentages (MPE) in lattice constant for different semiconductors. Data is from Ref. \citep{jana2018}}
	\label{fig_r_latcon}       
\end{figure}

Optimized structure is the basic criteria for further study on any material. Geometric optimization is done using energy minimization technique, where variation of energy is found by varying the lattice parameters of any solid (variable cell relaxation), as well as, by changing relative positions of atoms (ionic relaxation). Thus, the energy minimised structure seems to be most stable structure. Although LDA is a very basic approximation, it can predict the lattice constants quite well. This is because the HEG or its slowly varying approximation is suitable for solids, though for molecules it is not realistic. PBE GGA incorporates the slowly varying density idea to LDA, so, performs better than LDA (see Table \ref{table_latt-cons}). One can see from this table, MPE and MAPE are same for PBE, so, PBE consistently overestimates the lattice constants for semiconductor. Now, let us take the case of Ge. The deviation is quite high, i.e., PBE over-binds the atoms so much. This is because of the deviation of PBE from GEA result, as we have discussed. PBEsol is better suited for solids, and, the MPE and MAPE proves this simple correction is one of the best performer in the test on lattice constants, even in this modern generation of DFT functionals. First non-empirical MGGA, TPSS is built upon PBE, so, overestimation is also evident. However, the SCAN MGGA functional can reduce the error by appropriate constraint management, but, the consistency as seen in PBE or TPSS, is not that much robust in SCAN. TM reduces the error significantly, and, the consistency is manageable \citep{jana2018, mo2017}. In Fig. \ref{fig_r_latcon} the MPE for different semiconductors are plotted taking the values from \citep{jana2018}. 

\begin{table}
	\centering
	\caption{\label{table_latt-cons} Energy minimized lattice constants (in \AA)~ of different semiconductors calculate using Projected Augmented Wave (PAW) basis set. The mean relative error (MPE), and
		mean absolute relative error (MAPE) are reported  with respect to experimental values.  Here A4, B1 and B3 are for Diamond, Rocksalt and Zincblende structures, respectively. Data is from Ref. \citep{jana2018}. }
		\begin{tabular}{p{2cm}p{1.3cm}p{1.3cm}p{1.3cm}p{1.3cm}p{1.3cm}p{1.3cm}p{1cm}}
		\hline\noalign{\smallskip}
		Methods$\rightarrow$ Solids $\downarrow$ &{\bf LSDA} & {\bf PBE}& {\bf PBEsol}& {\bf TPSS}&  {\bf SCAN}& {\bf TM}& {\bf Expt.}\\
		\noalign{\smallskip}\svhline\noalign{\smallskip}
C (A4)                                    & 3.536  & 3.573 & 3.557  & 3.572 & 3.555 & 3.554 & 3.567  \\
Si (A4)                                   & 5.400  & 5.467 & 5.433  & 5.450 & 5.425 & 5.411 & 5.430  \\
Ge (A4)                                   & 5.648  & 5.785 & 5.704  & 5.754 & 5.687 & 5.672 & 5.652  \\
SiC (B3)                                  & 4.332  & 4.379 & 4.359  & 4.365 & 4.352 & 4.344 & 4.358  \\
BN (B3)                                   & 3.583  & 3.625 & 3.607  & 3.624 & 3.605 & 3.608 & 3.607  \\
BP (B3)                                   & 4.490  & 4.546 & 4.521  & 4.545 & 4.521 & 4.510 & 4.538  \\
BAs (B3)                                  & 4.742  & 4.817 & 4.778  & 4.810 & 4.779 & 4.763 & 4.777  \\
AlP (B3)                                  & 5.433  & 5.504 & 5.470  & 5.489 & 5.478 & 5.450 & 5.460  \\
AlAs (B3)                                 & 5.637  & 5.732 & 5.681  & 5.707 & 5.670 & 5.656 & 5.658  \\
AlSb (B3)                                 & 6.120  & 6.232 & 6.168  & 6.208 & 6.173 & 6.143 & 6.136  \\
$\beta-$GaN (B3)                          & 4.503  & 4.588 & 4.547  & 4.581 & 4.524 & 4.549 & 4.531  \\
GaP (B3)                                  & 5.425  & 5.533 & 5.474  & 5.523 & 5.457 & 5.464 & 5.448  \\
GaAs (B3)                                 & 5.627  & 5.763 & 5.684  & 5.737 & 5.664 & 5.664 & 5.648  \\
GaSb (B3)                                 & 6.067  & 6.226 & 6.130  & 6.190 & 6.117 & 6.102 & 6.096  \\
InP (B3)                                  & 5.878  & 6.001 & 5.932  & 5.989 & 5.938 & 5.923 & 5.866  \\
InAs (B3)                                 & 6.061  & 6.211 & 6.122  & 6.182 & 6.122 & 6.104 & 6.054  \\
InSb (B3)                                 & 6.472  & 6.651 & 6.543  & 6.611 & 6.545 & 6.521 & 6.479  \\
ZnS (B3)                                  & 5.403  & 5.440 & 5.355  & 5.401 & 5.370 & 5.364 & 5.409  \\
ZnSe (B3)                                 & 5.570  & 5.734 & 5.634  & 5.681 & 5.652 & 5.633 & 5.668  \\
ZnTe (B3)                                 & 5.995  & 6.178 & 6.064  & 6.115 & 6.077 & 6.056 & 6.089  \\
CdS (B3)                                  & 5.758  & 5.926 & 5.824  & 5.933 & 5.856 & 5.857 & 5.818  \\
CdSe (B3)                                 & 6.009  & 6.195 & 6.080  & 6.192 & 6.100 & 6.102 & 6.052  \\
CdTe (B3)                                 & 6.405  & 6.610 & 6.291  & 6.604 & 6.521 & 6.497 & 6.480  \\
MgO (B1)                                  & 4.145  & 4.242 & 4.206  & 4.224 & 4.184 & 4.202 & 4.207  \\
MgS (B3)                                  & 5.580  & 5.684 & 5.642  & 5.681 & 5.634 & 5.629 & 5.202  \\
MgSe (B1)                                 & 5.382  & 5.501 & 5.445  & 5.491 & 5.454 & 5.435 & 5.400  \\
MgTe (B3)                                 & 6.365  & 6.506 & 6.439  & 6.500 & 6.452 & 6.422 & 6.420  \\
CaS (B1)                                  & 5.570  & 5.710 & 5.632  & 5.698 & 5.683 & 5.657 & 5.689  \\
CaSe (B1)                                 & 5.798  & 5.955 & 5.869  & 5.947 & 5.921 & 5.894 & 5.916  \\
CaTe (B1)                                 & 6.215  & 6.389 & 6.291  & 6.386 & 6.375 & 6.317 & 6.348  \\
SrS (B1)                                  & 5.910  & 6.056 & 5.973  & 6.047 & 6.031 & 6.007 & 5.990  \\
SrSe (B1)                                 & 6.129  & 6.297 & 6.203  & 6.286 & 6.264 & 6.234 & 6.234  \\
SrTe (B1)                                 & 6.531  & 6.714 & 6.609  & 6.708 & 6.693 & 6.641 & 6.640  \\
BaS (B1)                                  & 6.289  & 6.433 & 6.362  & 6.448 & 6.441 & 6.390 & 6.389  \\
BaSe (B1)                                 & 6.510  & 6.681 & 6.577  & 6.670 & 6.659 & 6.622 & 6.595  \\
BaTe (B1)                                 & 6.890  & 7.080 & 6.964  & 7.075 & 7.071 & 7.012 & 7.007  \\ 
\hline\hline
MPE (\% )                                  & -0.678 & 1.502 & 0.186  & 1.238 & 0.556 & 0.271 & --     \\
MAPE (\% )                                 & 1.099  & 1.502 & 0.789  & 1.247 & 0.723 & 0.588 & --    \\    
		\noalign{\smallskip}\hline\noalign{\smallskip}
	\end{tabular} 
\end{table}

\subsubsection{Bulk Modulus}
\begin{figure}[b!]
	\centering
	\sidecaption
	\framebox{\includegraphics[scale=0.4]{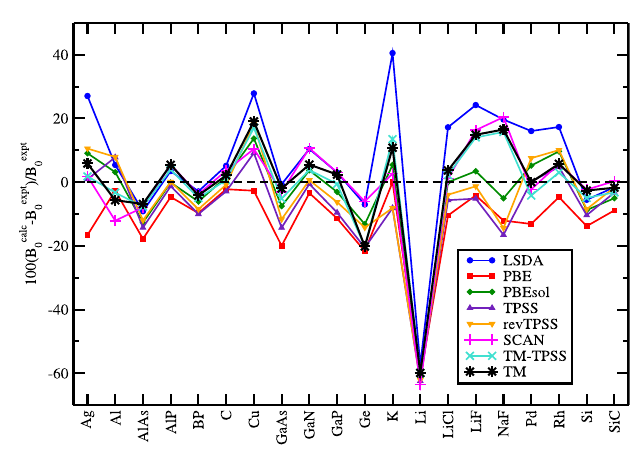}}
	\caption{Relative error percentage (MPE) in prediction of bulk moduli for different solids.  MPE(\%) and MAPE(\%) for
		LSDA$\rightarrow 9.752, 12.260$; PBE$\rightarrow -8.521, 9.136$; PBEsol$\rightarrow -0.331, 6.380$;
		TPSS$\rightarrow -4.617, 7.389$; revTPSS$\rightarrow -1.384, 7.518$; SCAN$\rightarrow ~~~~-1.797, 7.105$; 
		TM-TPSS$\rightarrow 2.748, 7.684$; TM$\rightarrow 4.435, 8.282$. Reproduced with permission from \citep{jana2018}.}
	\label{fig_r_strc}       
\end{figure}

The bulk modulus of any solid is the unit of measurement of its resistance towards external compression. It is the ratio of the infinitesimal pressure change, and, the fractional change of the volume. In DFT calculation, the definition is given in terms of energy as: $B=\mathscr{V} \frac{\partial^2 E}{\partial \mathscr{V}^2}$, and, found through an energy of state (EOS) equation fitting for the energy ($E$) vs. volume ($\mathscr{V}$) data. One of the widely used EOS is Birch-Murnaghan (BM) equation \citep{birch1947}:
\begin{equation}\label{eq_BM}
E(\mathscr{V}) = E_0 + \frac{9 B_0 \mathscr{V}_0}{16} \left\{ \left[ \left(\frac{\mathscr{V}_0}{\mathscr{V}}\right)^{2/3} -1  \right]^3 B_0' 
+ \left[ \left(\frac{\mathscr{V}_0}{\mathscr{V}}\right)^{2/3} -1  \right]^2 \left[ 6-4 \left(\frac{\mathscr{V}_0}{\mathscr{V}}\right)^{2/3}  \right]  \right\}
\end{equation}
Here, $\mathscr{V}_0$, $B_0$, and, $E_0$ are the equilibrium volume bulk modulus, and, the energy; $B_0'$ is the pressure gradient of $B_0$. After fitting the data into the curve, the equilibrium pressure can be calculated using $P(\mathscr{V}) = -\frac{d E(\mathscr{V})}{d \mathscr{V}}$.
There are several other EOS available for different pressure ranges (refer to \citep{angel2014}).

In the Fig. \ref{fig_r_strc}, we present the plot provided in Ref. \citep{jana2018}. The PBEsol is intended for solids, and, is always best performer in GGA segment. \tcr{Even it performs  better than most of the MGGAs}. The TM-TPSS scheme \citep{jana2018}, in which the TM exchange is mixed with TPSS correlation, performs best among the semi-local functionals but considering both the lattice constant and bulk modulus calculation, may be, PBEsol is the best choice. Here, we should mention that vLB-FP-NMTO, which incorporates a potential only exchange correction to LDA, have provided excellent lattice constant and bulk moduli matchings for C$_3$N$_4$ polymorphs \citep{datta2020carbon}. Also for group IV and III-V semiconductors it has performed well \citep{datta2019}. This is an important finding, as, \tcr{finding out } a semi-local correction which performs good for both structural, and, electronic structural calculations is still an open challenge. 

\subsubsection{Transition Pressure for Structural Phase Transition}
\begin{table}[b!]
	\caption{\label{table_phase-tran} Transition pressure of different semiconductors in GPa. Data is taken from Ref. \citep{sengupta2018} for EXX and RPA, and, from Ref. \citep{shahi2018} for others, including experimental values as presented. Experimental pressure at right side of $|$ is for forward transition and at left side for reverse transition.Crystal structures: A4$\equiv$Diamond; A5$\equiv$$\beta$-Sn; B1$\equiv$Rocksalt; B3$\equiv$Zincblende; B4$\equiv$Wurtzite; B8$_1$$\equiv$Nickeline; B9$\equiv$Cinnabar; B33$\equiv$Cncn.  }
	\begin{tabular}{p{3cm}p{1cm}p{1cm}p{1cm}p{1cm}p{1cm}p{2.5cm}}
		\hline\noalign{\smallskip}
		{Methods $\rightarrow$} & {\textbf{LDA}} & {\textbf{PBE}} & {\textbf{SCAN}} & \textbf{EXX} & \textbf{RPA} & {\textbf{Expt.}} \\
		\noalign{\smallskip}\svhline\noalign{\smallskip}
		Si (A4 $\rightarrow$  A5) & 7.1 & 9.8 & 14.5 & 51.4 & 13.8 & 11.3 - 12.6 \\
		Ge (A4 $\rightarrow$  A5) & 6.6 & 7.9 & 11.3 & 51.1 & 11.2 & 10.6(5) \\
		SiC (B3  $\rightarrow$  B1) & 59.4 & 64.8 & 74.0 & 114.6 & 74.3 & 35|100 \\
		GaAs (B3  $\rightarrow$  B33) & 12.4 & 14.1 & 17.1 & 60.2 & 18.9 & 11.2|17.3 \\
		GaP (B3  $\rightarrow$  B33) & 18.1 & 20.7 & 25.9 & -- & -- & 26 \\
		GaN (B4  $\rightarrow$  B1) & 42.3 & 46.2 & 42.1 & -- &--  & 30|47; 52.2; 37 \\
		InN (B4  $\rightarrow$  B1) & 8.9 & 12.2 & 10.6 & -- & -- & 12.1; 10 \\
		AlN (B4  $\rightarrow$  B1) & 7.2 & 13.1 & 12.5 & -- & -- & 0|14; 22; 20 \\
		InP (B3  $\rightarrow$  B1) & 6.2 & 8.4 & 10.6 & -- & -- & 9.8(5); 10.8(5) \\
		InAs (B3  $\rightarrow$  B1) & 4.2 & 6.0 & 7.5 & -- &--  & 7.0 \\
		AlP (B3  $\rightarrow$  B8$_1$) & 6.8 & 9.4 & 11.5 & -- & -- & 4.8|14.2 \\
		AlAs (B3  $\rightarrow$  B8$_1$) & 6.7 & 8.9 & 10.7 & -- & -- & 2|12 \\
		AlSb (B3  $\rightarrow$  B33) & 3.7 & 5.1 & 6.6 & -- & -- & 2.2|8.1 \\
		ZnO (B4  $\rightarrow$  B1) & 9.1 & 11.6 & 8.8 & -- & -- & 1.9(2)|9.1(2);  9.8 \\
		ZnS (B3  $\rightarrow$  B1) & 15.2 & 16.8 & 18.3 & -- & -- & 10|14.7; 17.4; 16.9 \\
		ZnSe (B3  $\rightarrow$  B1) & 12.1 & 13.7 & 15.9 & -- & -- & 12 - 20 \\
		ZnTe (B3  $\rightarrow$  B9) & 8.7 & 9.7 & 10.5 & -- & -- & 8|9.5 \\
		CdS (B4  $\rightarrow$  B1) & 2.4 & 4.4 & 2.9 & -- & -- & 1.2|2.54; 3 \\
		CdSe (B4  $\rightarrow$  B1) & 2.4 & 4.1 & 3.3 & -- & -- & 1.7|2.72 \\
		CdTe (B3  $\rightarrow$  B9) & 3.8 & 4.8 & 4.2 & -- & -- & 2.67|3.53; 3.8 \\
		HgS (B9  $\rightarrow$  B1) & 11.1 & 15.8 & 21.9 &--  & -- & 20.5(7) \\
		HgSe (B9  $\rightarrow$  B1) & 6.7 & 10.4 & 15.8 & -- & -- & 14.6(6); 15.5 \\
		HgTe (B9  $\rightarrow$  B1) & 2.3 & 5.2 & 6.6 & -- & -- & 8.0 \\
		\noalign{\smallskip}\hline\noalign{\smallskip}
	\end{tabular}
\end{table}

Polymorphism is common among materials, where a material having a particular chemical formula can be found in different structures having different space-group symmetries. Under external influence, e.g, temperature and pressure, these materials can switch from one symmetrical structure to another. Determination of the critical pressure at which such structural phase transition takes place is not as simple as determining lattice constants, and, presents a tough test for an approximate density functional.

There are some techniques for critical pressure determination, one of the easiest is through the Gibbs energy matching. The Gibbs free energy is expressed as $\mathcal{G}(\mathscr{V},T)= E(\mathscr{V},T)-T~\mathcal{S}(\mathscr{V},T) + P(\mathscr{V},T)~ \mathscr{V}$, which reduces to enthalpy $\mathcal{H}(\mathscr{V})=E(\mathscr{V}) +P(\mathscr{V})~ \mathscr{V}$ at zero temperature ($\mathcal{S}$ is entropy). Now, at constant pressure and temperature, the Gibbs energy per formula unit of the two phases must be equal at phase equilibrium condition. At zero temperature, the equilibrium pressure for two different phases are found using the EOS fit, and, the transition pressure is found by an iterative solution starting from an initial guess, aiming towards the matching of enthalpies \citep{sengupta2018}.

The structural phase transition pressure for a set of materials calculated by Sengupta \etal for zero temperature and room temperature ($300K$) have shown that the SCAN GGA tends to be the most accurate (semi-)local functional which yields comparable results with RPA and experiments \citep{sengupta2018}. Sahin \etal have found similar trend (see, Table \ref{table_phase-tran}).

\subsection{Electronic Properties}
We have seen that the PBEsol GGA, or, SCAN and TM MGGA are good enough in structural property prediction. In electronic structural case, the situation is not same. It is a tough job for non-empirical semi-local functionals to reproduce the exact band structural properties, band gaps, and, band alignments. The reason is well discussed \citep{cohen2012} through the absence of derivative discontinuity or self-interaction error, delocalization error. We have discussed the theoretical background of these terms in previous sections. 

There habe been a common belief that, the KS potentials does not possess any derivative discontinuity.
Kraisler and Kronik have shown this is indeed not true, and, generally all xc functionals possess a non-zero $\Delta v_{xc}$ \citep{kraisler2014} originated through the ensemble averaging for fractional charge within KS scheme. For small finite systems, addition of $\Delta v_{xc}$, even in simple LDA, significantly improve the calculated bandgap, however, for bulk systems (infinitely large systems tending solid state limit), the gap deduced from the total energy difference reduces to the KS gap. The reason they have pointed that while HO and LU orbitals are delocalized, the xc kernel in (semi-)local cases are very much local. Different approaches have been proposed, either by localizing HO and LU orbitals through dielectric screening \citep{chan2010}, or, by imposing SIC \citep{pederson1988}. An alternative way is to impose the asymptotic behaviour within a solid locally, which can be implemented  easily in site-centred basis-set methods like LMTO-ASA or FP-NMTO \citep{NMTO}.

\begin{figure}[h!]
	\centering
	\sidecaption
	\framebox{\includegraphics[scale=0.19]{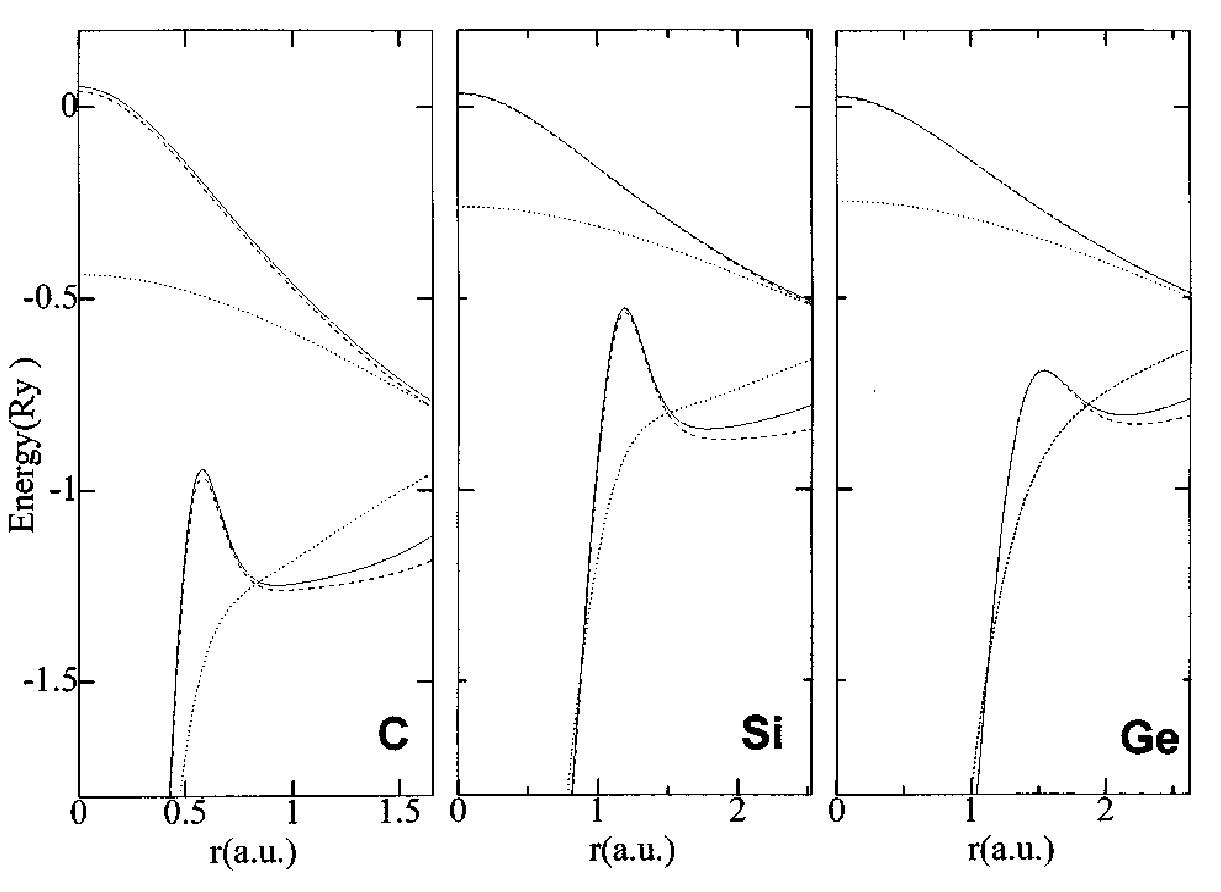}}
	{\includegraphics[scale=0.19]{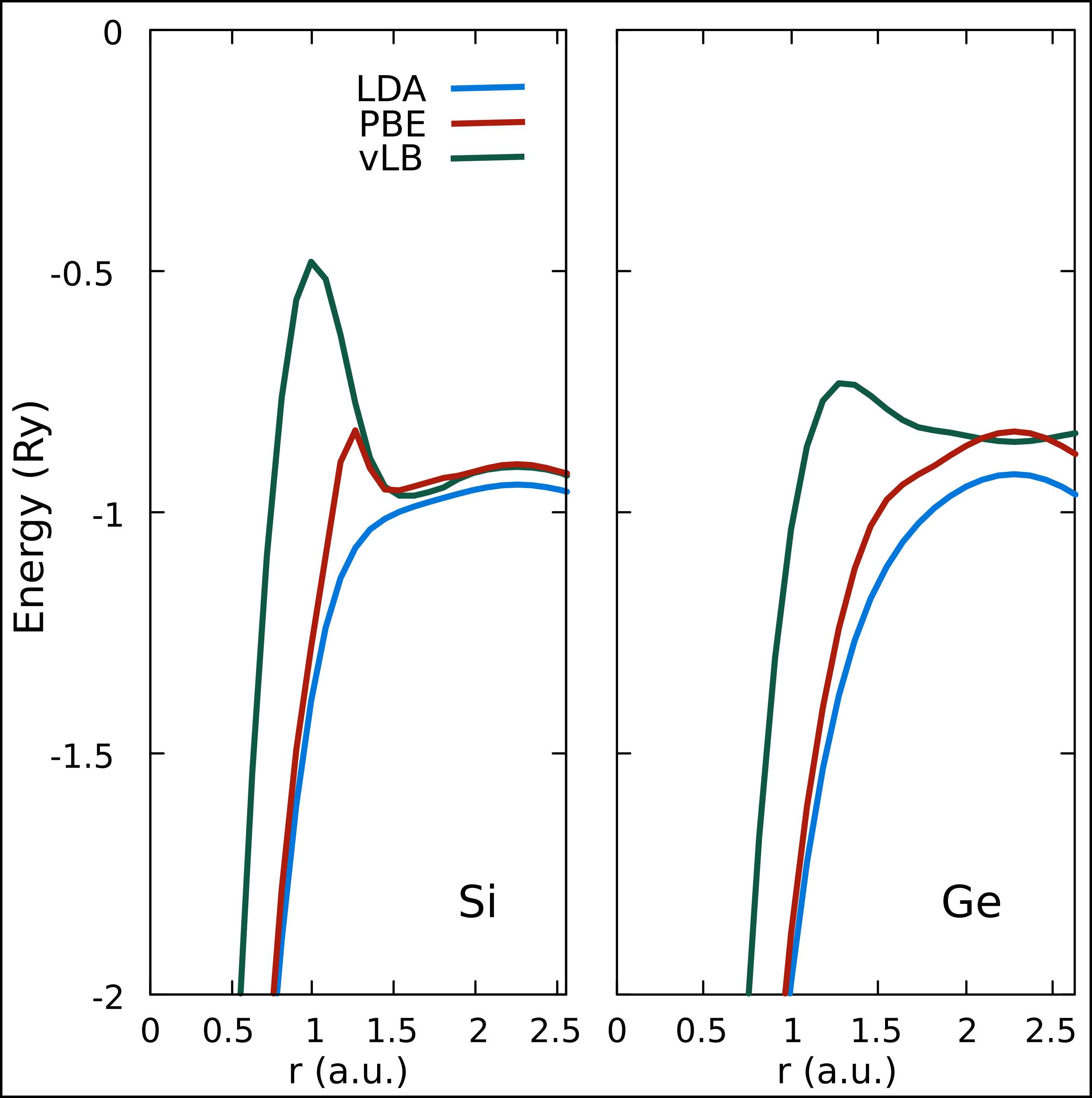}}
	\caption{(left) Exact exchange (EXX) potential (lines with kink) and LDA potential (smooth lines) for C, Si, Ge calculated using KKR-ASA and LMTO-ASA methods. (Reproduced with permission from \citep{kotani1996}). (right) LDA, PBE and vLB xc potentials for Si and Ge calculated using self-consistent FP-NMTO method \citep{datta2019}.}
	\label{fig_r_xcpot}       
\end{figure}

We have talked so much about exchange-correlation potentials. How does it look like? Kotani \etal have presented the exact exchange (EXX) potentials for different solids in a series of work, calculated by OPM method using LMTO-ASA and KKR-ASA (Fig. \ref{fig_r_xcpot}) \citep{kotani1995, kotani1996}. These work as the benchmark for semi-empirical GGAs and MGGAs. As seen in right panel of Fig. \ref{fig_r_xcpot}, the vLB-NMTO potential for Si and Ge can reproduce the EXX nature. Another thing to observe is the shell-closure like behaviour of the PBE and vLB-NMTO potentials, i.e, the potentials are not smooth like LDA, but, having \tcr{similar} kinks as observed in EXX calculation (left panel). 

\subsubsection{Band Gaps}

\begin{figure}[t!]
	\centering
	\sidecaption
	{\includegraphics[scale=0.3]{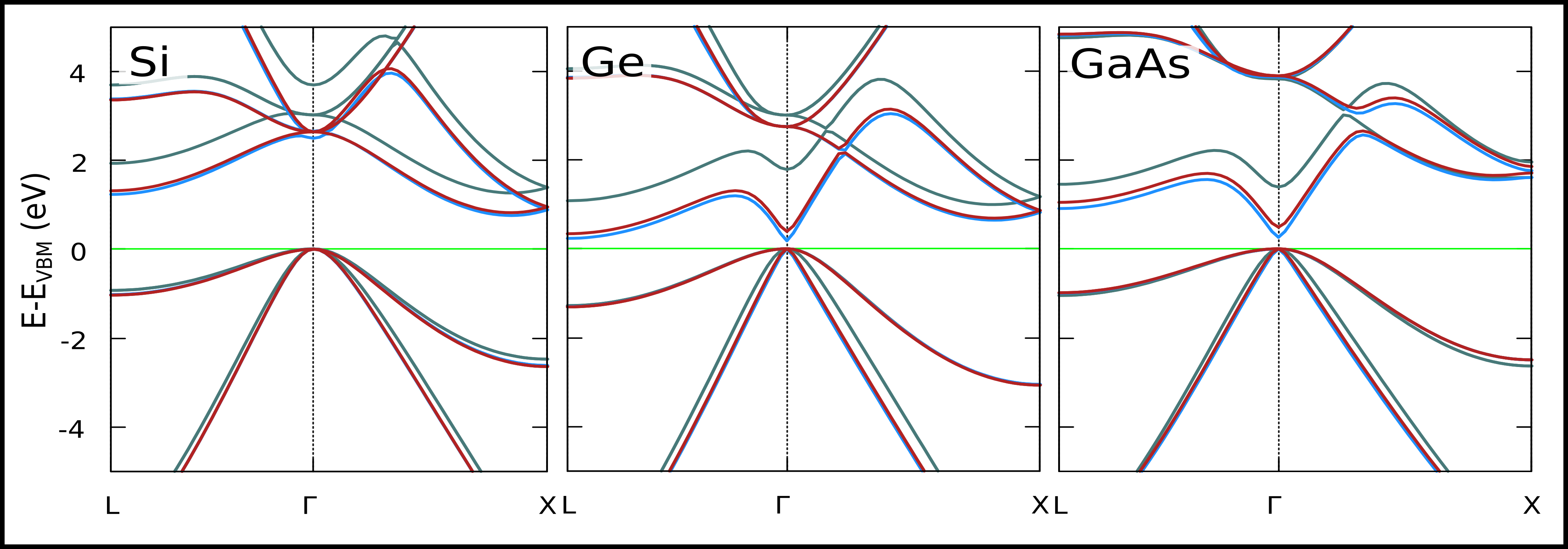}}
	\caption{Self-consistent FP-NMTO band structure (L$-\Gamma-$X) of Si, Ge and GaAs for LDA (blue), PBE (red), and LDA+vLB (green) DFT approximations \citep{datta2019}.}
	\label{fig_r_bands}
\end{figure}

Bandgap prediction is the most stringent test for DFT methods and the non empirical (semi-)local functionals still fail here. The underestimation can be as large as $80-90\%$ in some cases. Let us take the example of Ge; most of the (semi-) local functionals predict it as zero or nearly zero gap semiconductors. The PBE value for Ge using PAW basis gives exactly zero value \citep{jana2018}. But what we have already mentioned the values may differ from one method to another, using PBE the self consistent FP-NMTO method predicts $0.33eV$, and, FP-LAPW (linear augmented plane wave) based Wien2k predicts $0.06eV$ gap of Ge \citep{tran2018}. The LDA and PBE predict Si and Ge as direct bandgap semiconductors, which is fundamentally wrong. In Fig. \ref{fig_r_bands} the bands for Si, Ge and GaAs using PBE, LDA and vLB-NMTO are presented for better understanding of the problem.

\begin{table}[b]
	\centering
	\caption{Bandgaps for different group IV (diamond structure) and group III-V (zincblende structure) semiconductors compared with experimental result. The AK13, GLLB-SC and TBmBJ values are form \citep{tran2018} (FP-LAPW basis set within Wien2k), PBE, TPSS, SCAN, TM values from \citep{jana2018} (PAW basis set within VASP) and all other values are from \citep{datta2019} (LMTO-ASA and FP-NMTO).}
	\label{table_bg}
	\begin{tabular}{p{1.8cm}p{1.2cm}p{1.2cm}p{0.8cm}p{0.8cm}p{0.8cm}p{0.8cm}p{0.8cm}p{1.0cm}p{0.9cm}}
		\hline\noalign{\smallskip}
		& \textbf{Si} & \textbf{Ge} & \textbf{GaP} & \textbf{GaSb} & \textbf{GaAs} & \textbf{InAs} &  \textbf{InP} &\textbf{MPE} & \textbf{MAPE} \\
		\noalign{\smallskip}\svhline\noalign{\smallskip}
		LDA(LMTO) & 0.49 & 0.10 & 1.67 & 0.52 & 0.08 & 0.01 & 0.6 & -65.51 & 65.51 \\
		vLB(LMTO) & 1.21 & 0.06 & 1.46 & 0.05 & 0.04 & 0.01 & 0.5 & -68.46 & 69.43 \\
		LDA(NMTO) & 0.79 & 0.19 & 1.53 & 0.25 & 0.33 & 0.01 & 0.5 & -64.39 & 64.39 \\
		PBE(NMTO) & 0.85 & 0.33 & 1.68 & 0.51 & 0.54 & 0.04 & 0.90 & -48.45 & 48.45 \\
		vLB(NMTO) & 1.25 & 0.86 & 1.87 & 0.94 & 1.43 & 0.39 & 1.18 & -1.77 & 12.95 \\
		PBE & 0.64 & 0.00 & 1.51 & 0.00 & 0.15 & 0.00 & 0.37 & -77.75 & 77.75 \\
		TPSS & 0.67 & 0.00 & 1.72 & 0.00 & 0.38 & 0.00 & 0.54 & -72.22 & 72.22 \\
		SCAN & 0.85 & 0.06 & 1.89 & 0.12 & 0.80 & 0.00 & 0.87 & -58.44 & 58.44 \\
		TM & 0.56 & 0.29 & 1.56 & 0.46 & 0.84 & 0.00 & 0.73 & -54.61 & 54.61 \\
		AK13 & 1.60 & 0.70 & 2.60 & 0.76 & 1.45 & 0.73 & 1.81 & 18.55 & 23.18 \\
		GLLB-SC & 1.06 & 0.24 & 2.56 & 0.32 & 1.05 & 0.07 & 1.51 & -33.63 & 38.40 \\
		TBmBJ & 1.15 & 0.83 & 2.25 & 0.95 & 1.64 & 0.67 & 1.62 & 14.64 & 15.99 \\
		HSE06 & 1.28 & 0.82 & 2.47 & 0.72 & 1.21 & 0.39 & 1.64 & 0.19 & 11.85 \\
		GW & 1.29-1.31 & 0.65-0.71 & 2.80 & 0.62 & 1.58 & 0.31 & 1.44 & -3.19 & 13.80 \\
		EXX & 1.50 & 1.01 & -- & -- & 1.82 & -- & -- &  &  \\
		\hline
		\textbf{Expt.} &  \textbf{1.17} &  \textbf{0.74} &  \textbf{2.32} &  \textbf{0.81} &  \textbf{1.52} &  \textbf{0.43} &    \textbf{1.42} \\
		\noalign{\smallskip}\hline\noalign{\smallskip}
	\end{tabular}
\end{table}

\begin{figure}[]
	\centering
	\sidecaption
	\fbox{\includegraphics[scale=0.47]{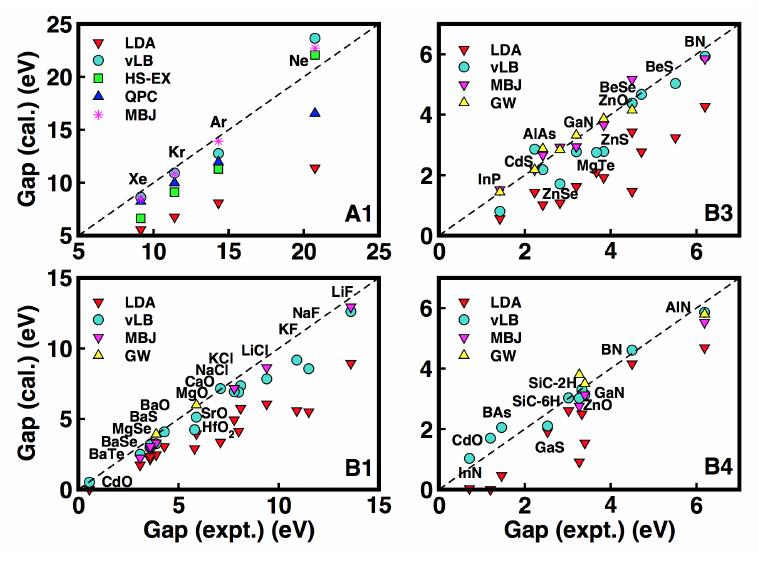}}
	\caption{Bandgaps for materials with four structures: {\bf (Left)} {\bf A1} (FCC) with $\beta = 0.04 - 0.082$, {\bf B1} (Rocksalt) with $\beta = 0.04 - 0.08$, {\bf (right)} {\bf B3} (Zincblende) with $\beta = 0.03 - 0.075$, {\bf B4} (Wurtzite) with $\beta = 0.03 - 0.09$ calculated using vLB-IP (denoted by VLB) compared with LDA, TBmBJ (denoted by MBJ) and GW methods. Reproduced with permission from \citep{singh2016}.}
	\label{fig_r_vLB-IP}
\end{figure}
In Table \ref{table_bg} we have presented data of bandgaps for  group IV and group III-V semiconductors. Larger list can be found in some articles, e.g., in Ref. \citep{tran2018}. For this group of materials, bandgaps are highly underestimated using PBE, TPSS, SCAN, TM, GLLB-SC and vLB-LMTO functionals, whereas, AK13 and TBmBJ overestimate. The HSE06 hybrid is the most successful, and, the most consistent performer in bandgap prediction. The vLB-NMTO is the best performing (semi-)local functional for this particular group of orthodox semiconductors followed by TBmBJ. Both of these are simple to implement in localized basis sets FP-NMTO and FP-LAPW, respectively. Intuitively we can say, the full potential description of the potential in interstitial region is the key of success \tcr{for these functionals}. This can be understood when we compare the gaps calculated by vLB-LMTO-ASA and vLB-FP-NMTO, the FP description is proved to be superior than the ASA. The IP-vLB-LMTO, which is not a single shot calculation, is comparable to these (see, Fig.\ref{fig_r_vLB-IP}). Though for this group of materials, GLLB-SC as implemented within FP-LAPW-Wien2k can not perform well, for larger variety of materials, MPE and MAPE for GLLB-SC is much lower \citep{tran2019}. For such diverse semiconducting materials the performance of TBmBJ is excellent. So, we can conclude that if a proper (semi-)local approximations are chosen \tcr{along with most compatible basis sets} (as TBmBJ-FPLAPW and vLB-NMTO), the bandgap prediction can be as good as calculated from computationally costly screened hybrid functional or Many Body Perturbation Theory (MBPT). \footnote[2]{MBPT: Due to the Coulomb repulsion between the electrons, around any electron a depletion of negative charge forms. This positive screening charge termed as electron-hole, and, the electron itself produce a quasiparticle. Quasiparticles interacts weakly with each-other via a screened Coulomb potential. The mathematical description involves the single-particle Green function $G(\vr,\vr',E)$ and quasiparticle self energy due to screened Coulomb interaction $W(\vr,\vr',E)$. For quasiparticle excitations, GW approximation with different levels of complexities are applied to find the properties involving excited state \citep{aulbur2000}.}


Oxide semiconductors are the most difficult set of solids to manage through DFT schemes due to their strong electronic correlation. For oxide semiconductors, performance of different hybrid functionals for prediction of both bandgap and dielectric constants can be found in Ref. \citep{he2017}. They have shown that (Table \ref{tab_oxides}) in both predictions, the HSE06 screened hybrid performs better than PBE0 and its self consistent version scPBE0. Though the betterment is quite prominent in bandgap prediction, in case of dielectric constant calculation, there is not so much improvement. 

\begin{svgraybox}
	\textbf{\large B4: Two-Dimensional Systems}
	
	Following the discovery of two dimensional (2D) carbon allotrope graphene, the research on 2D material become prolific \citep{novoselov2004}. A plethora of new 2D materials have been proposed theoretically, and, a few of those are also experimentally synthesized \citep{chowdhury2016,bandyopadhyay2020}. In theoretical prediction on 2D materials, the periodicity along the perpendicular direction of the layer have to be broken, but, mostly the DFT packages rely on the Bloch theorem imposing periodic boundary condition in all three direction. So, to mimic the situation for 2D systems, a large vacuum have to be introduced. The main difficulty comes from the long-range Coulomb interaction, which may lead to long-ranged monopolar or dipolar image interactions that fall off rather slowly with periodic separation \citep{makov1995}. As a result truncated Coulomb interaction for 2D systems is necessary \citep{ismail2006}. On the other hand, LDA or GGA do not take into account the long-range, non-local correlations, whereas, for layered systems long-ranged van der Waals (vdW) interaction is important. So, for non-monolayer layered systems, vdW correction have been introduced through approximations. A good description and performance comparison can be found in Ref. \citep{berland2015,lebedeva2017}.
	
	However, standard DFT methods perform within acceptable accuracy. Beyond graphene, 2D allotropes of group IV elements (silicene, germannene) \citep{vogt2012,davila2014,csahin2009}, group V elements (phosphorene, arsenene, antimonene, bismuthene) \citep{shao2018,reis2017,shah2020,vierimaa2016,ersan2016,kadioglu2018} or III-V binary systems have shown coherence between theoretical and experimental observations through the matching of structural and electronic properties \citep{cahangirov2009}. Even the level of buckling in some 2D honeycomb systems (e.g, silicene, germanene, arsenene, etc.) have been well predicted. As a consequence, theoretical prediction of the properties of 2D systems, even of 1D linear chains rely on the orthodox DFT methods \citep{ataca2008,ersan2020,ersan2019}. 
\end{svgraybox}

\subsubsection{Accurate Determination of Fermi Energy, Work Function and Band Alignment}

The work-function (WF) of any piece of material is the energy required to remove an electron from its surface. In $0K$, electrons can occupy the states upto the Fermi energy ($E_F$) level of the material. The energy at a point outside the surface is the vacuum potential energy ($E_{vac}$). Work function is the difference of these two: $WF= E_{vac} - E_F$.
In intrinsic semiconductors, $E_F$ lies at the middle of VBM and CBM, so, in absolute scale, the alignment of the energy bands can be detected accordingly. This makes the $E_F$ calculation so important \citep{west2012} along with the bandgap estimation.

\begin{figure}[t]
	\centering
	\sidecaption
	\includegraphics[scale=0.35]{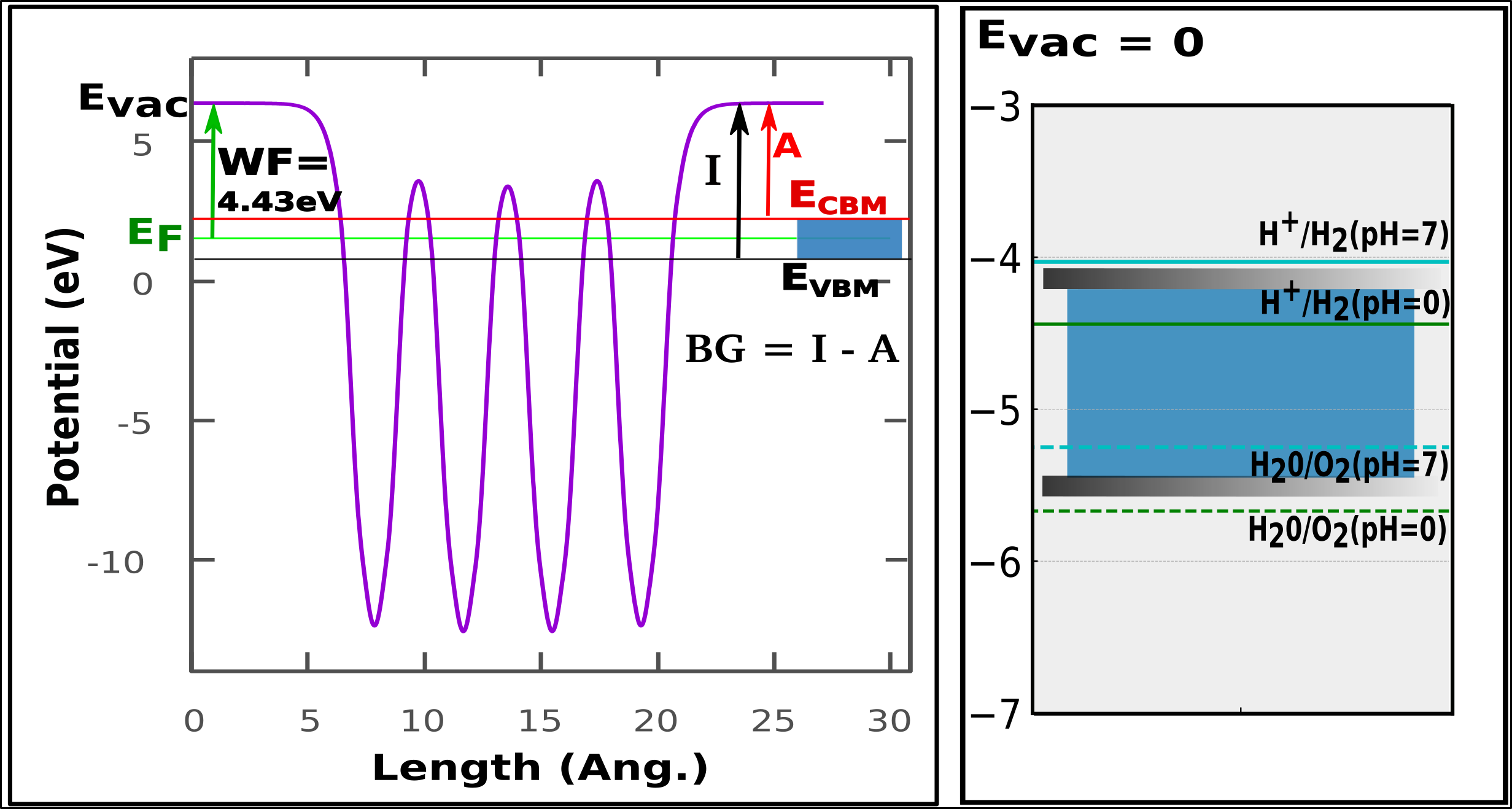}
	\caption{(left)HSE06  calculated work-function of Si slab (4.43 eV) is in good agreement with experimental value 4.87 eV \cite{hollinger1983}. (right)The resulting CBM-VBM alignment plot. I$\equiv$Ionization energy, A$\equiv$Electron affinity, BG$\equiv$bandgap.}
	\label{fig_r_WF-Si}
\end{figure}

To mimic aperiodicity in the direction perpendicular to the material surface, a large vacuum have to be introduced. Two different calculations are carried out, one on the bulk system, and, another on a slab made by some number of layers. The electrostatic potential within the interstitial region of the slab and the bulk should be same. The difference of the macroscopic averages over the plane parallel to the surface-plane for bulk and for slab ($\Delta E_{es} = E_{bulk} - E_{slab}$) works as a constant correctional shift to $E_F^{Bulk}$ to bring equivalence of energy levels of these two separate self-consistent calculations \citep{wf_theo}, and finally, we find:
\begin{equation}
WF = E_{vac} - E_{F,corrected}^{Bulk} ~~\text{where,}~~~
E_{F,corrected}^{Bulk} = E_F^{Bulk} - \Delta E_{es}
\end{equation}

It is shown earlier that the PBE can not detect the Fermi energy accurately for semiconductors \citep{west2012}. As an example, we show the work function and band alignment of Si using HSE06 hybrid functional. 
The calculated work-function using HSE06, $4.43 eV$ for Si is in very good agreement with experimental value $4.87 eV$ \cite{hollinger1983}. Similar trend is observed for 2D-graphene (HSE06$\rightarrow 4.39eV$; expt. $\rightarrow4.56$) \citep{datta2020pccp}. 
Semiconductor band alignment analysis is done for a group of traditional semiconductors in Ref. \citep{hinuma2014} and HSE06 has performed great in locating so. 

The proper determination of band alignment has come more into focus in the modern era of energy efficiency and green energy hunt.  To initiate the water-redox reaction in semiconductors, the band-edge (CBM and VBM) positions are equally important as its bandgap. In normal hydrogen electrode (NHE) scale, the VBM must be more positive than the water oxidation level ($E_{H_2O/O_2}$=1.23, 0.81 V for pH=0, 7), and, the CBM must be more negative than the Hydrogen production energy ($E_{H^+/H_2}$=0, $-0.41$ V for pH=0, 7 vs. NHE). There are many photocatalytic candidates, and in many instances, the DFT calculations can accurately predict the VBM and CBM positions of those materials. As an example, we mention matching of the graphitic \cn values using HSE06. For this reason, HSE06 is widely used for theoretical prediction of new photocatalytic materials \citep{datta2020carbon, li2017Review}. 

\begin{figure}[h!]
	\centering
	\sidecaption
	{\includegraphics[scale=0.28]{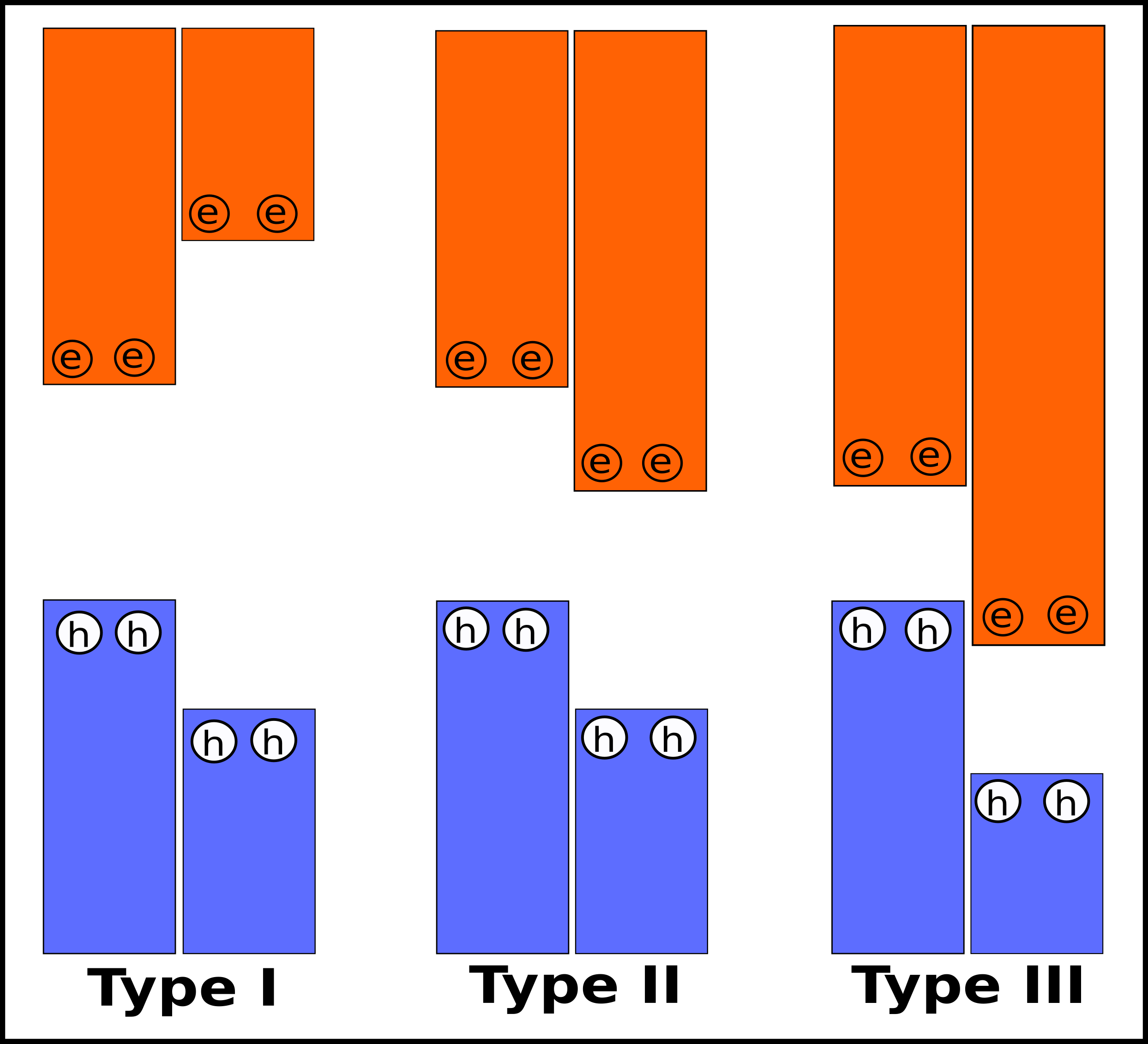}}
	\caption{Band alignments for Type I, II and III semiconductor heterostructures with electron-hole (e-h) indicated.}
	\label{fig_r_hetero}       
\end{figure}

Another useful application of band alignment prediction is in formation of different heterostructures. Semiconductor heterostructures are of three types Type I, II and III, as indicated in the Fig. \ref{fig_r_hetero}. Type I semiconductors are useful in optical devices (lasers, LEDs, etc.), Type II are used in efficient photocatalytic material designing, high electron mobility transistors, and, Type III in tunnelling field effect transistors.  
Band offset of semiconductor alloy heterostructure which is important for different barrier formations is studied for a set of materials by Wadehra \etal. They have shown the efficiency of HSE06 hybrid  calculation \citep{wadehra2010} (See, Table \ref{tab_hetero}).

\begin{table}[]
	\centering
	\caption{\label{tab_hetero} Comparison of HSE06 and PBE calculated VBM and CBM offsets (eV) of III-V heterostructures with experiment values $\Delta E_{\textrm{v/c}} = E^{S2}_{\textrm{v/c}} -E^{S1}_{\textrm{v/c}} $. All the alloys are in the form A$_{0.5}$B$_{0.5}$C. Asterisks indicate that HSE06 with $w_{SR}^{HSE} = 0.30$ instead of the default $w_{SR}^{HSE} = 0.25$. Opposite sign of VBM and CBM offset indicates Type-I		and the same sign indicates Type-II heterostructures. Reproduced with permission from \citep{wadehra2010}.}
	\begin{tabular}{p{2cm}p{0.9cm}p{0.7cm}p{1.0cm}p{0.9cm}p{0.7cm}p{0.7cm}}
		\hline\noalign{\smallskip}
		\multicolumn{1}{l}{Heterostructure} & \multicolumn{3}{l}{$\Delta E_{\text{v}}$ (eV)} & \multicolumn{3}{l}{$\Delta E_{\text{c}}$ (eV)} \\
		S1/S2 & HSE06 & PBE & Exp. & HSE06 & PBE & Exp. \\
				 \noalign{\smallskip}\svhline\noalign{\smallskip}
		AlAs/GaAs & 0.52* & 0.45 & 0.53 & -1.02* & -1.08 & -1.05 \\
		AlP/GaP & 0.54 & 0.49 & 0.55 & 0.58 & 0.57 & 0.38 \\
		AlSb/GaSb & 0.38 & 0.35 & 0.38 & -0.67 & -1.21 & -0.51 \\
		AlGaAs/GaAs & 0.26* & 0.21 & 0.27 & -0.42* & -1.12 & -0.31 \\
		InGaP/GaAs & 0.32 & 0.26 & 0.31 & -0.24 & -0.34 & -0.18 \\
		InP/InGaAs & 0.36 & 0.27 & 0.34 & -0.38 & -0.42 & -0.27 \\
		InP/AlInAs & 0.16 & 0.14 & 0.17 & 0.22 & 0.17 & 0.25 \\
		AlInP/InGaP & 0.22 & 0.19 & 0.24 & -0.23 & -0.74 & -0.26 \\
		AlInAs/InGaAs & 0.23 & 0.18 & 0.22 & -0.57 & -0.54 & -0.51 \\
		InGaP/AlGaAs & 0.11 & 0.08 & 0.09 & 0.25 & 0.19 & 0.28 \\
		d/o InGaP & 0.01 & 0.02 &  & -0.24 & -0.18 & 0.15\\
		\noalign{\smallskip}\hline\noalign{\smallskip}
		\end{tabular}
\end{table}

\begin{table}[]
	\centering
	\caption{ \label{tab_oxides} Bandgaps and Dielectric Constants of different oxides. Data for PBE0, HSE06 and Experiments are from Ref. \citep{he2017} (PAW-VASP) and PBE, AK13, TBmBJ and GLLB-SC are from \citep{tran2018} (FPLAPW-Wien2k) }
	\begin{tabular}{p{1.2cm}p{1.2cm}p{0.6cm}p{0.7cm}p{0.9cm}p{0.9cm}p{0.7cm}p{0.8cm}p{1.0cm}p{0.7cm}p{0.8cm}p{0.6cm}}
		\hline\noalign{\smallskip}
		&   & \multicolumn{7}{l}{\textbf{Bandgap (eV)}} & \multicolumn{3}{l}{\textbf{Dielec. Const.}}  \\
		Solids & Spc. grp. & PBE & AK13 & TBmBJ & GLLB-SC & PBE0 & HSE06 & \textbf{Expt.} & PBE0 & HSE06 & \textbf{Expt.}  \\
		\noalign{\smallskip}\svhline\noalign{\smallskip}
		MgO & Fm$\bar{3}$m & 4.78 & 6.69 & 7.13 & 8.31 & 7.20 & 6.48 & 7.9 & 2.88 & 2.91 & 2.94   \\
		ZnO & P6$_3$mc & 0.81 & 2.06 & 2.65 & 2.57 & 3.15 & 2.49 & 3.44 & 3.67 & 3.74 & 3.72   \\
		BeO & P6$_3$mc & 7.65 & 9.42 & 9.66 & 11.36 & 10.25 & 9.52 & 10.59 & 2.92 & 2.93 & 2.95   \\
		SnO$_2$ & P42/mnm & 1.24 & 2.20 & 3.19 & 3.33 & 3.40 & 2.71 & 3.6 & 3.23 & 3.92 & 4.06   \\
		SiO$_2$ & P3$_1$21 & 5.93 & 8.17 & 8.70 & 9.88 & 8.41 & 7.67 & 9.0 & 2.32 & 2.32 & 2.3   \\
		TiO$_2$ & P4$_2$/mnm & 1.89 & 2.23 & 2.56 & 3.77 & 3.86 & 3.12 & 3.05 & 6.41 & 6.45 & 7.37   \\
		Cu$_2$O & Pn$\bar{3}$m & 0.53 & 0.84 & 0.81 & 1.10 & 2.80 & 2.07 & 2.17 & 6.28 & 6.34 & 7.11   \\
		Al$_2$O$_3$ & R$\bar{3}$c & 6.26 & 7.92 & 8.34 & 9.82 & 8.72 & 7.99 & 8.8 & 3.03 & 3.03 & 3.4   \\
		CdO & Fm$\bar{3}$m &  &  &  &  & 1.52 & 0.88 & 0.84 & 5.04 & 5.12 & 6.2   \\
		PbO & P4/nmm &  &  &  &  & 2.57 & 1.93 & 2.03 & 5.32 & 5.55 & 7.1   \\
		GeO$_2$ & P4$_2$/mnm &  &  &  &  & 4.26 & 3.56 & 5.35 & 3.83 & 3.80 & 4.43   \\
		HfO$_2$ & P4$_2$/nmc &  &  &  &  & 7.07 & 6.34 & 5.9 & 4.39 & 4.39 &    \\
		Ag$_2$O & Pn$\bar{3}$m &  &  &  &  & 1.90 & 1.21 & 1.20 & 5.34 & 5.70 &    \\
		La$_2$O$_3$ & P$\bar{3}$m1 &  &  &  &  & 6.14 & 3.92 & 5.8 & 4.02 & 4.05 &    \\
		In$_2$O$_3$ & R$\bar{3}$c &  &  &  &  & 3.38 & 2.72 & 3.02 & 3.92 & 3.68 & 3.62   \\
		CuAlO$_2$ & R$\bar{3}$m &  &  &  &  & 4.24 & 3.49 & 2.99 & 4.37 & 4.42 &    \\
		LiCoO$_2$ & R$\bar{3}$m &  &  &  &  & 4.89 & 4.12 & 2.7 & 4.63 & 4.64 &    \\
		LaAlO$_3$ & R$\bar{3}$c &  &  &  &  & 6.30 & 5.57 & 6.33 & 4.06 & 4.05 & 4.0   \\
		LiNbO$_3$ & R3c &  &  &  &  & 5.71 & 4.97 & 3.50 & 4.32 & 4.41 & 4.87   \\
		BiFeO$_3$ & R3c &  &  &  &  & 4.14 & 3.40 & 2.67 & 6.10 & 6.12 & 5.52   \\
		BaTiO$_3$ & P4mm &  &  &  &  & 3.77 & 3.05 & 3.26 & 5.52 & 5.54 & 5.75   \\
		PbTiO$_3$ & P4mm &  &  &  &  & 3.32 & 2.62 & 3.4 & 6.43 & 6.58 & 6.25   \\
		BaSnO$_3$ & Pm$\bar{3}$m &  &  &  &  & 3.14 & 2.45 & 3.1 & 3.98 & 4.25 & 3.3   \\
		SrTiO$_3$ & Pm$\bar{3}$m &  &  &  &  & 3.85 & 3.33 & 3.3 & 5.52 & 5.35 & 6.1   \\
		LaMnO$_3$ & Pmna &  &  &  &  & 3.01 & 2.27 & 1.7 & 5.11 & 5.16 & 4.9   \\
		BiVO$_4$ & C2/c &  &  &  &  & 3.67 & 2.97 & 2.4 & 6.34 & 6.41 &    \\
		Ag$_2$PdO$_2$ & Immm &  &  &  &  & 1.89 & 1.19 & 0.18 & 7.34 & 7.39 &    \\
		BiCuSeO & P4/nmm &  &  &  &  & 1.92 & 1.31 & 0.8 & 10.18 & 10.46 &    \\
		LaCuSeO & P4/nmm &  &  &  &  & 3.36 & 2.68 & 2.8 & 6.36 & 6.43 &    \\
		\noalign{\smallskip}\hline\noalign{\smallskip}
		\multicolumn{2}{l}{\textbf{MARE  (\%)}}  &  &  &  &  & 31.1 & 18.5 &  & 9.7 & 8.8 &  \\
		\noalign{\smallskip}\hline\noalign{\smallskip} 
	\end{tabular}
\end{table}

\subsection{Optical Properties}
Optical properties of matter are directly dependent to its electronic structure, as, inter and intra band transitions of electrons are responsible for that. The minimum amount of photon energy required to create a hole in VB, and, lifting a electron to CB in any semiconductor should be greater than or equal to its bandgap. Photons with energy higher than the bandgap initiate transitions to higher energy levels of conduction band. So, proper production of energy bands becomes so much important.

In DFT scheme, the dielectric response is calculated using perturbative method, easily through RPA, or, through more complicated time dependent DFT (TDDFT) calculations (may refer Chap. 4 of \citep{fiolhais2003} for TDDFT). A DFT formulation of RPA for dielectric response is done in Ref. \citep{hybertsen1987}.

The complex dielectric tensor $\tilde{d}_{\alpha\beta}(\omega)$ can be defined as:
\begin{align}
\tilde{d}_{\alpha\beta}(\omega) =\tilde{d}_{\alpha\beta}^{(r)}+ {\bf i} \tilde{d}^{(i)}_{\alpha\beta}
&= 1+\frac{4 \pi e^2}{\Omega N_{\textbf{k}} m^2}\sum\limits_{n,n'}\sum\limits_{\textbf{k}}
\frac{\langle u_{\textbf{k},n'}\vert\hat{\textbf{p}}_{\alpha}\vert u_{\textbf{k},n}\rangle 
	\langle u_{\textbf{k},n}\vert\hat{\textbf{p}}_{\beta}^{\dagger} \vert u_{\textbf{k},n'}\rangle}
{(E_{\textbf{k},n'}-E_{\textbf{k},n})^2} \times \nonumber\\
&\left[\frac{f(E_{\textbf{k},n})}{E_{\textbf{k},n'}-E_{\textbf{k},n}+\hbar\omega+ {\bf i}\hbar\Gamma} +
\frac{f(E_{\textbf{k},n})}{E_{\textbf{k},n'}-E_{\textbf{k},n}-\hbar\omega-{\bf i} \hbar\Gamma}\right] 
\end{align}\label{eq_dielec}
The lifetime of electrons in excited states should not be infinite. To retain a finite lifetime of excited-states, a small positive value of inter-band broadening parameter $\Gamma$ is introduced to produce an intrinsic broadening. The imaginary ($\Im$) part of dielectric tensor, $\tilde{d}^i_{\alpha\beta}$ is found first and the real ($\Re$) part $\tilde{d}^r_{\alpha\beta}$ is calculated using Kramers-Kronig relation: 
$\tilde{d}^r_{\alpha \beta}(\omega)=1+\frac{2}{\pi}\int_{0}^{\infty}\frac{\omega' \tilde{d}^i_{\alpha \beta}(\omega')} {\omega'^{2}-\omega^{2}}d\omega'$. These are used to calculate optical conductivity, refractive index (RI) and absorption coefficient.
\begin{align}
&\text{Opt. Cond.:  } \Re [\sigma_{\alpha\beta} (\omega)]= \frac{\omega}{4\pi}\tilde{d}^{(i)}_{\alpha\beta}(\omega) 
~;~ \text{Abs. Coeff.: } A_{\alpha\alpha}(\omega)=\frac{2\omega n^-_{\alpha\alpha}(\omega)}{c} \nonumber\\ 
&\text{Complex RI:  } \tilde{n}_{\alpha\alpha}=\tilde{n}^+_{\alpha\alpha}+{\bf i} \tilde{n}^-_{\alpha\alpha} 
~\text{where, } n^{\pm}_{\alpha\alpha}(\omega)= \sqrt{\frac{|\tilde{d}_{\alpha\alpha}(\omega)| \pm \tilde{d}^{(r)}_{\alpha\alpha}(\omega)}{2}} 
\end{align}\label{eq_optical}

\begin{figure}[b!]
	\centering
	\sidecaption
	\framebox{\includegraphics[scale=0.4]{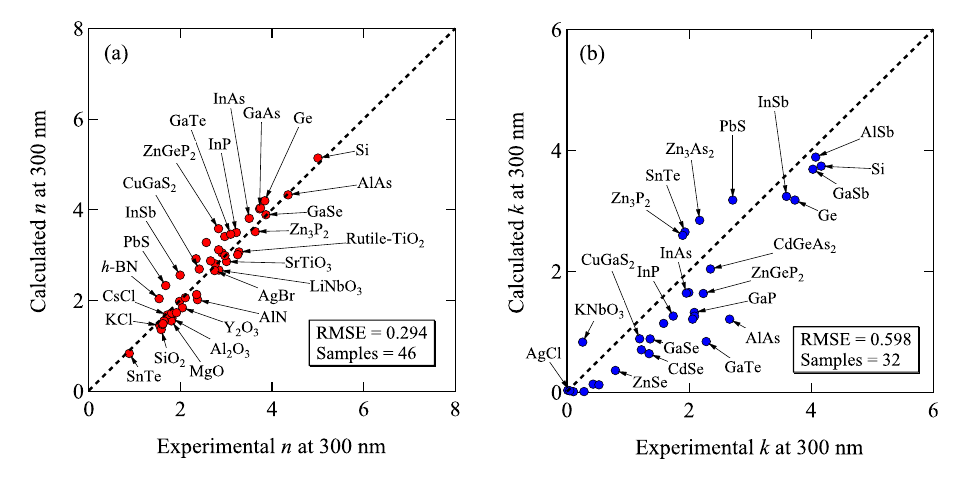}}
	\framebox{\includegraphics[scale=0.4]{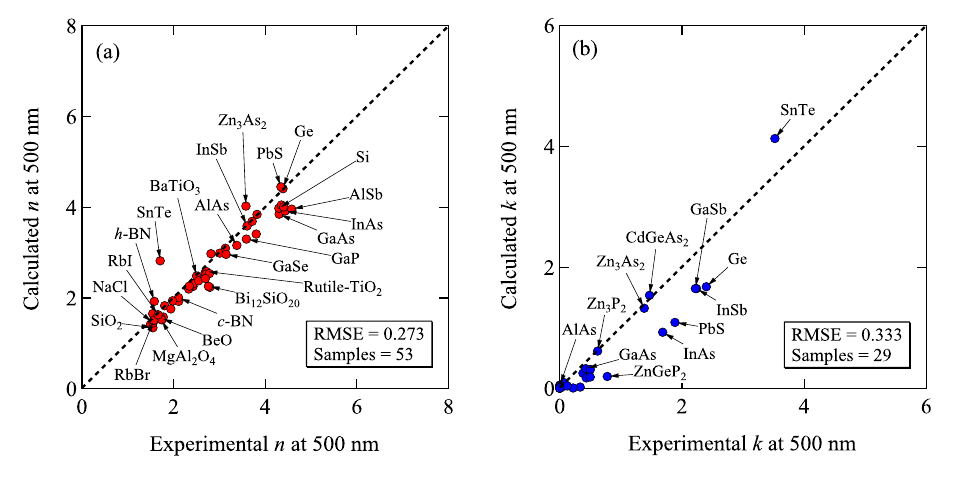}}
	\caption{Comparison of components of complex refractive index $\tilde{n}$ calculated using TBmBJ with experimental values for $300 nm$ and $500 nm$. In figure, $\tilde{n}^+ \equiv n$ and $\tilde{n}^- \equiv k$. Reproduced with permission from \citep{nakano2018}.}
	\label{fig_r_optical}       
\end{figure}

No doubt, which approximation can reproduce the exact band structure should be mostly successful in optical property calculations. Now, the CB belongs to the excited state, and, that is why MBPT is successful in producing the optical response well. However, the computational cost of MBPT is much higher than DFT based calculations. DFT based approximations free from bandgap underestimation \tcr{problem} are often used in optical property calculation successfully \citep{nishiwaki2020,sabetvand2020,ersan2019}.
For oxides a comparison of the performance of dielectric constants calculated using different hybrid functionals is done by He \etal \citep{he2017} (see, Table \ref{tab_oxides}). TBmBJ MGGA  predicted optical constants, as well as, the bandgaps agree well with experimental observations. The root mean squared error (RMSE) for the calculated values is within $0.440 eV$ for bandgap, $0.246-0.299$ and $0.207-0.598$ for $\tilde{n}^+$ and $\tilde{n}^-$ components of complex refractive index  \citep{nakano2018}. A comparison can be visualised in Fig. \ref{fig_r_optical}.
These small RMSE values project the effectiveness of TBmBJ  in predicting the optical property of semiconductors in the ultraviolet to visible range of spectrum.

\subsection{Magnetic Properties}
\begin{table}[b!]
	\centering
	\caption{\label{table_MM} Spin magnetic moment per unit cell of elementary  ferromagnetic transition metals, and, atomic spin magnetic moment of antiferromagnetic transition metal-oxides in $\mu_{\text{B}}$/atom unit. SCAN and SCAN-L values are form \citep{mejia2019}, TM \& TM-TPSS values are from \citep{jana2018} and others are from \citep{tran2018}.}
	\begin{tabular}{p{1.4cm}p{1.2cm}p{1.2cm}p{1.4cm}p{1cm}p{1.2cm}p{1.2cm}p{1.2cm}p{0.7cm}}
		\hline\noalign{\smallskip}
		\textbf{Methods} & \textbf{Fe}    & \textbf{Co}         & \textbf{Ni}  & \textbf{MnO}  & \textbf{FeO}  & \textbf{CoO}  & \textbf{NiO}  & \textbf{CuO}  \\ 
		\noalign{\smallskip}\svhline\noalign{\smallskip}
		{LDA}     & 2.21           & 1.59                & 0.61   & 4.11 & 3.33 & 2.36 & 1.21 & 0.12     \\
		{PBE}     & 2.22           & 1.62                & 0.64   & 4.17 & 3.39 & 2.43 & 1.38 & 0.38     \\
		{vLB}     & 2.02           & 1.39                & 0.41   & 3.93 & 3.02 & 1.75 & 0.67 & 0.00     \\
		{AK13}    & 2.58           & 1.70                & 0.69   & 4.39 & 3.51 & 2.59 & 1.57 & 0.54     \\
		{GLLB-SC} & 3.08           & 1.98                & 0.81   & 4.56 & 3.74 & 2.73 & 1.65 & 0.55     \\
		{BJ}      & 2.39           & 1.63                & 0.62   & 4.19 & 3.40 & 2.48 & 1.48 & 0.50     \\
		{TBmBJ}   & 2.51           & 1.69                & 0.73   & 4.41 & 3.58 & 2.71 & 1.75 & 0.74     \\
		{TPSS}	  & 2.19		   & 1.61				 & 0.63	  &--&--&--&--& --   \\
		{SCAN}    & 2.60           & 1.80                & 0.78    &--&--&--&--& --   \\
		{SCAN-L}  & 2.05           & 1.63                & 0.67    &--&--&--&--&   -- \\
		{TM	} 	  & 2.22		   & 1.60				 & 0.60	  &--&--&--&--& --   \\
		{TM-TPSS} & 2.25		   & 1.64				 & 0.69	  &--&--&--&--& --   \\				
		{HSE06}   & 2.79           & 1.90                & 0.88   & 4.36 & 3.55 & 2.65 & 1.68 & 0.67     \\
		{Expt.}   & 1.98-2.08      & 1.52-1.62     & 0.52,0.55    & 4.58 & 3.32-4.6 &3.35-3.98 &1.9,2.2 & 0.65\\ 
		\noalign{\smallskip}\hline\noalign{\smallskip}
	\end{tabular}
\end{table}

Magnetic property determination involves spin dependent density functional approximation, which is available for almost all functionals in literature. In elemental $3d$ transition metal ferromagnets (Fe, Co, and Ni), LDA and PBE estimated magnetic moments are quite satisfying, PBE overestimates slightly. This is why PBE often successfully predicts the magnetic properties of advanced alloys as well \citep{mahat2020,mahat2021}.  The vLB as implemented using FP-LAPW underestimates, whereas, AK13, BJ, TBmBJ overestimates; GLLB-SC and HSE06 hugely overestimates.
Among (semi-)local functionals, SCAN functional is most successful in structural and mechanical property estimation. Surprisingly, it significantly overestimates the magnetization of elemental $3d$-transition metals. Mej{\'\i}a \etal have concluded in their study that it is originating from the insensitivity of the switching function to $\tilde{\alpha}$ in some particular range, and, oversensitivity in another range \citep{mejia2019}. \tcr{They have} proposed a revised version of SCAN, called SCAN-L.

Situation for transition metal-oxides is different. Tran \etal have compared the atomic spin magnetic moments for different oxides as well as of these elementary magnets \citep{tran2018}. Within FPLAPW scheme, LDA and vLB predicted atomic spin magnetic moments  of oxides are highly underestimated in most of the cases. All corrections over LDA improve the value, specially AK13. GLLB-SC, TBmBJ and HSE06 perform better than others.

\section{Conclusion}
Success of any theory is justified by its application. DFT is not an exception to that. In this chapter, starting from the very basic Hartree-Fock theory, we have showcased the development till recent time in DFT methodology, and, have compared the performance of different DFT approximations, and, have tried to provide a prescription which approximation should be used for which material property prediction. Sometimes, it becomes hard for the material scientists to grab all the DFT advances, but, have to apply those methods or approximations. This chapter is aimed to bridge this gap. There are so many intermediate researches that led to this development but could not be acknowledged in this chapter to make it compact. All of those are equally important.

Density functional theoretical development is a long journey, and, only a part of it is travelled till date. We have tried to sum up the story within a small volume, so that, those who are willing to join the material science theoretical and/or computational research can get a roadmap. This chapter must not be thought as an alternative of the wonderfully written books on DFT (e.g., Ref. \citep{ dreizler2012, sahni2016, parr1980, engel2013, fiolhais2003} etc.), but, should be regarded as an introduction to the beautiful field of DFT to motivate students for further extensive study.

\section*{Acknowledgements}
The authors gratefully recall the effort of \emph{(Late) Prof. Abhijit Mookerjee} in crafting the collaboration between the developers of methodology and material physicists. S. Datta expresses his gratitude towards \emph{Prof. Manoj Kumar Harbola}, who has introduced the beauty of DFT to him.

\begin{svgraybox}
\center{\underline{\textbf{Some Useful Symbols}}}\\
\begin{tabular}{p{5.5cm}|p{5.5cm}}
$\hat{H} \rightarrow$ Hamiltonion Operator & $\hat{T}$ \& $\hat{V} \rightarrow$	Kin. \& Pot. energy Operator \\
$\Psi({\bf X}) \rightarrow$ Spin dependent wavefn. & $\Phi({\bf R}) \rightarrow$ Spin independent wavefn. \\
$\Psi^S({\bf X}) \rightarrow$ Slater ,, & $\Phi^S({\bf R}) \rightarrow$ Slater ,,\\
$\psi_i(\vx_n)=\phi_i(\vr_n)\chi_i(\sigma_n) \rightarrow$ $i$-th Spin-orbital state at $\vr_n$ and of spin $\sigma_n$
& $\phi_i(\vr_n),\chi_i(\sigma_n) \rightarrow$ $i$-th Space-only orbital at $\vr_n$ and $i$-th spin state.\\
$\rho(\vr),\rzr \rightarrow$ Density and Ground State Density &
$\rho_1(\vr, \vr'),\rho_2(\vr_1\vr_2,\vr_1'\vr_2') \rightarrow$ 1st and 2nd order Spinless Reduced Density Matrix.\\
$\rho_{x}(\vr, \vr') \rightarrow$ Fermi Exchange Hole & $\rho_{xc}(\vr, \vr') \rightarrow$ Fermi-Coulomb xc Hole\\
$g_{xc} \rightarrow$ Pair-Correlation Function & $\mu_\pm \rightarrow$ Chemical Potential\\
$I^{(N)}\rightarrow$ Ionization pot. of $N$ $e$ system. & $A^{(N)}\rightarrow$ Electron Affinity.\\
$E \rightarrow$ Total Energy; $E_g \rightarrow$ Bandgap & $E_{ee}\rightarrow$ \emph{e-e} Interaction Energy\\
$E_H \rightarrow$ Hartree Energy & $E_{xc}\rightarrow$ Exchange-Correlation Energy\\
$\epsilon_i \rightarrow$ $i$-th orbital energy & $\varepsilon_{xc} \rightarrow$ xc Energy per particle.\\
$\tau(\vr) \rightarrow$ Kinetic Energy Density & $\tilde{f}\rightarrow$ Exchange Enhancement Factor. \\
$v_{eff}, v_{ext} \rightarrow$ Effective and External Pot. & $v_H, v_{xc} \rightarrow$ Hartree and xc Pot.
\end{tabular}
\end{svgraybox}


\bibliographystyle{spphys.bst}
\end{document}